\documentclass[cernpreprint, atlasdraft=false, UKenglish, texmf, orcidlogo]{atlasdoc}
\usepackage{atlaspackage}
\usepackage{atlasbiblatex}

\usepackage{multirow}

\usepackage{atlasphysics}
\graphicspath{{logos/}{figures/}}

\usepackage{ANA-TOPQ-2022-24-PAPER-defs}

\AtlasTitle{Measurement of the top quark mass with the ATLAS detector using $t\bar{t}$~events with a high transverse momentum top quark}

\AtlasAbstract{%
The mass of the top quark is measured using top-quark-top-antiquark pair events with high transverse momentum top quarks. The dataset, collected with the ATLAS detector in proton--proton collisions at $\sqrt{s}=13$~\ensuremath{\text{Te\kern-0.1em V}}\xspace delivered by the Large Hadron Collider, corresponds to an integrated luminosity of 140~\mbox{fb\(^{-1}\)}\xspace. The analysis targets events in the lepton-plus-jets decay channel, with an electron or muon from a semi-leptonically decaying top quark and a hadronically decaying top quark that is sufficiently energetic to be reconstructed as a single large-radius jet. The mean of the invariant mass of the reconstructed large-radius jet provides the sensitivity to the top quark mass and is simultaneously fitted with two additional observables to reduce the impact of the systematic uncertainties. The top quark mass is measured to be $m_t = 172.95 \pm 0.53$~\ensuremath{\text{Ge\kern-0.1em V}}\xspace, which is the most precise ATLAS measurement from a single channel.
}

\AtlasRefCode{TOPQ-2022-24}

\PreprintIdNumber{CERN-EP-2025-015}

\AtlasJournal{Phys.\ Lett.\ B.}
\AtlasJournalRef{\PLB 867 (2025) 139608}
\AtlasDOI{10.1016/j.physletb.2025.139608}

\hypersetup{pdftitle={ATLAS document},pdfauthor={The ATLAS Collaboration}}

\begin{document}

\maketitle

\tableofcontents

%

\section{Introduction}

The top quark is the heaviest fundamental particle observed to date and precise knowledge of its mass (\mt)
is crucial to test the consistency of the Standard Model (SM) of particle physics~\cite{ALEPH:2010aa, Degrassi:2012ry, Bezrukov:2007ep, Bezrukov:2012sa, DeSimone:2008ei}.
Since the discovery of the top quark at the Tevatron~\cite{CDF:1995wbb, D0:1995jca}, the CDF and D0\ collaborations have made multiple measurements of \mt, culminating in the
2016 combined result~\cite{TeV:2016combination}.
The two general-purpose experiments at the Large Hadron Collider (LHC), ATLAS~\cite{PERF-2007-01} and CMS~\cite{CMS:Detector-2008}, produced
multiple measurements of \mt using data collected in proton--proton ($pp$) collisions at $\sqrt{s}=7$~\TeV~and $8$~\TeV.
The combination of these measurements is $\mt = 172.52 \pm 0.33~\GeV$~\cite{TOPQ-2019-13}.
The large dataset provided by the LHC during collisions at $\sqrt{s}=13$~\TeV corresponding to an integrated luminosity of 140~\ifb increases substantially the number of top quarks available for measurements.
Measurements of \mt performed with a partial $13$~\TeV dataset~\cite{CMS-TOP-17-008, CMS-TOP-17-001, CMS-TOP-19-005, CMS-TOP-19-009,  CMS-TOP-20-008,  CMS-TOP-21-008, TOPQ-2017-17} include the most precise single measurement to date of $171.77 \pm 0.37$~\GeV~\cite{CMS-TOP-20-008} from the CMS Collaboration.
The $13$~\TeV dataset also opens the possibility to make measurements in rare phase-space regions, for example events where the top quarks have high transverse momenta (\pt).
In this boosted regime, the top-quark decay products are collimated and are often captured in a single large-radius (large-$R$) jet.
This simplifies the reconstruction of hadronically decaying top quarks compared with the inclusive phase space, potentially reducing the systematic uncertainties in \mt.
In this regime the CMS Collaboration measured $ \mt = 173.06 \pm 0.84$~\GeV~\cite{CMS-TOP-21-012}.

In this Letter, the first ATLAS measurement of the top quark mass using boosted top quarks is presented. The measurement targets events consistent with top quark pair production
in the lepton-plus-jets decay channel ($\ttbar\to W^+W^-b\bar{b} \to \ell\nu  q\bar{q}^\prime b\bar{b}$~with~$\ell = e \text{~or~} \mu$)
and the hadronically decaying top quark is reconstructed using a single large-$R$~jet ($R=1.0$).
The large-$R$ jets used in this Letter are formed by reclustering $R=0.4$~jets~\cite{Nachman_2015} such that the energy and mass scales of the large-$R$~jets are largely determined by the precise jet energy scale of the $R=0.4$~jets~\cite{JETM-2018-05}.
The selection and reconstruction applied largely follow those of the ATLAS cross-section measurement in the same channel~\cite{TOPQ-2019-23}.
The mean of the invariant mass of the selected large-$R$~jet is the observable sensitive to \mt and two additional observables are defined to control and reduce the impact of the systematic uncertainties.
One observable is sensitive to the jet energy scale and the other is sensitive to radiation from the $b$-quarks that originate from the decay of top quarks.
The top quark mass is obtained from a profile-likelihood fit to the three observables in which the systematic uncertainties are included as nuisance parameters.
The combination of the boosted selection, improved understanding of the flavour dependence of the jet energy response~\cite{STDM-2020-04,ATL-PHYS-PUB-2022-021} and the fit strategy lead to smaller systematic uncertainties than in previous ATLAS \mt measurements.

The fit model in the measurement uses Monte Carlo (MC) simulation that is based on a next-to-leading-order~(NLO) calculation in quantum chromodynamics (QCD) of \ttbar production.
This is interfaced to a parton shower (PS) algorithm that provides resummation of soft and collinear QCD radiation and a non-pertubative hadronisation model that simulates the formation of hadrons.
The top quark mass is a renormalisation-scheme-dependent parameter in perturbative quantum field theory. The precise identification of the \mt parameter in MC simulations within a field-theoretic mass scheme is the subject of theoretical studies~\cite{Hoang:2018zrp, Azzi:2019yne, Hoang:2020iah, Dehnadi:2023msm}. The boosted top-quark regime may, in the future, offer the possibility to connect to hadron-level calculations where the top quark mass is unambiguously defined~\cite{Hoang:2020iah, Dehnadi:2023msm}.


\section{Data and simulation}
\label{sec:dataMC}

The data were recorded with the ATLAS detector at the Large Hadron Collider during $pp$~collisions at $\sqrt{s}=13$~\TeV
using triggers designed to select events containing high \pt~electrons or muons~\cite{TRIG-2018-01,TRIG-2018-05,TRIG-2019-03}.
The integrated luminosity of the dataset is 140~\ifb~\cite{DAPR-2021-01}.
The ATLAS experiment~\cite{PERF-2007-01} is a multipurpose particle detector
with a forward--backward symmetric cylindrical geometry and a near \(4\pi\) coverage in
solid angle.\footnote{\AtlasCoordFootnote}
It consists of an inner tracking detector surrounded by a thin superconducting solenoid
providing a \qty{2}{\tesla} axial magnetic field, electromagnetic and hadronic calorimeters, and a muon spectrometer.
The inner tracking detector covers the pseudorapidity range \(|\eta| < 2.5\).
It consists of silicon pixel, silicon microstrip, and transition radiation tracking detectors.
Lead/liquid-argon (LAr) sampling calorimeters provide electromagnetic (EM) energy measurements
with high granularity within the region \(|\eta|< 3.2\).
A steel/scintillator-tile hadronic calorimeter covers the central pseudorapidity range (\(|\eta| < 1.7\)).
The endcap and forward regions are instrumented with LAr calorimeters
for EM and hadronic energy measurements up to \(|\eta| = 4.9\).
The muon spectrometer surrounds the calorimeters and is based on
three large superconducting air-core toroidal magnets with eight coils each.
The field integral of the toroids ranges between \num{2.0} and \qty{6.0}{\tesla\metre}
across most of the detector.
The muon spectrometer includes a system of precision tracking chambers up to \(|\eta| = 2.7\) and fast detectors for triggering up to \(|\eta| = 2.4\).
The luminosity is measured mainly by the LUCID--2~\cite{LUCID2} detector, which is located close to the beampipe.
A two-level trigger system is used to select events~\cite{TRIG-2016-01}.
The first-level trigger is implemented in hardware and uses a subset of the detector information
to accept events at a rate close to \qty{100}{\kHz}.
This is followed by a software-based trigger that
reduces the accepted rate of complete events to \qty{1.25}{\kHz} on average
depending on the data-taking conditions.
A software suite~\cite{SOFT-2022-02} is used in data simulation, in the reconstruction
and analysis of real and simulated data, in detector operations, and in the trigger and data acquisition
systems of the experiment.

Monte Carlo simulated event samples are used in the analysis to model the \ttbar\ events and most of the background processes.
Most samples are processed through a full simulation of the ATLAS detector~\cite{SOFT-2010-01} based on \GEANT~\cite{Agostinelli:2002hh}.
A few samples used for evaluating the systematic uncertainties employ a faster simulation setup that makes use of parameterised showers in the calorimeters~\cite{ATL-PHYS-PUB-2010-013}.
All simulated samples are overlaid with additional $pp$~interactions (pileup), generated with \textsc{Pythia}~8.186~\cite{Sjostrand:2007gs} using the \textsc{NNPDF2.3lo}~\cite{Ball:2012cx} set of parton distribution functions (PDF) and the A3 set of tuned parameters~\cite{ATL-PHYS-PUB-2016-017}. The average number of interactions per bunch crossing is reweighted to match that in data.
Events are processed with the same reconstruction pipeline as the data. A \mt value of $\SI{172.5}{\GeV}$ is used in all samples, unless stated otherwise.

The production of \ttbar events is modelled using the \POWHEGBOX[v2]~\cite{Frixione:2007nw,Nason:2004rx,Frixione:2007vw,Alioli:2010xd} generator, which provides matrix elements (ME) at NLO in QCD in the production process (implemented via \POWHEG-hvq), and the \NNPDF[3.0nlo]~\cite{Ball:2014uwa} PDF set. The \hdamp parameter, which effectively regulates the high transverse momentum~(\pT) radiation against which the \ttbar system recoils, is set to 1.5\mt~\cite{ATL-PHYS-PUB-2016-020}.
The renormalisation (\muR) and factorisation (\muF) scales are dynamic, using the functional form: $\sqrt{\mt^{2} + p^2_{\mathrm{T}}(t)}$.
The events are interfaced with \PYTHIA[8.230]~\cite{Sjostrand:2014zea} for the parton shower and hadronisation, using the A14 set of tuned parameters~\cite{ATL-PHYS-PUB-2014-021} and the \NNPDF[2.3lo] set of PDFs. The value of the strong coupling constant in the final state shower of \PYTHIA ($\alpha_{s}^{\mathrm{FSR}}$) is 0.127. Matrix element corrections that approximate NLO QCD are enabled in \Pythia for all emissions, compensating for the leading-order (LO) precision used in \POWHEG-hvq to simulate the top quark decay. The \pthard parameter, which affects the matching of the parton shower to the ME calculation, is set to 0. The recoil parameter, which decides the recoil target for secondary gluon emissions from the \bquark in the $t\to Wb$ vertex, is set to the \bquark.\footnote{This corresponds to setting \texttt{recoilToColor=on} in the \texttt{TimeShower} class in \PYTHIA.} The early resonance decay parameter, controlling whether resonance decays can occur before or after colour reconnection (CR) happens in the simulation, is set such that resonance decays can only occur after CR\@.
Samples are generated with the full ATLAS detector simulation and five different \mt values: $171, 172, 172.5, 173$~and $174$~\GeV. The sample with $\mt=172.5$~\GeV is referred to as the nominal
$\ttbar$~sample.
An additional sample using the fast detector simulation is available for comparisons made to fast-simulation samples used for evaluating some of the systematic uncertainties.

Alternative \ttbar\ samples are generated to assess the systematic uncertainties in the modelling of \ttbar production.
Several samples make just one change in the MC settings compared to the nominal \ttbar sample, allowing
individual systematic effects to be investigated separately.
To probe uncertainties in final-state radiation (FSR), two samples are generated with alternative values of $\alpha_s^{\mathrm{FSR}}$, equal to 0.115 for the down variation and 0.142 for the up variation, corresponding to variations of the scale $\mu_{\mathrm{r}}^{\mathrm{FSR}}$ by a factor of two and a half.
Another sample is produced with the \hdamp parameter doubled to $3\mt$.
The ambiguities in matching the ME calculation and the parton shower are tested with a sample where the \pthard parameter is set to 1~\cite{ATL-PHYS-PUB-2023-029}.
The ambiguity in the choice of the recoil particle for the secondary gluon emission from a $b$~quark produced in $t\to Wb$~is addressed by generating a sample where the top quark takes part in the recoil (recoil-to-top).
Uncertainties in the modelling of the underlying event are evaluated with two samples where the settings for the $\alpha_s$ value used in the multiple parton interactions (MPI) and
the CR range of the proton beam remnants are changed according to the VAR1 eigentune of \Pythia~\cite{ATL-PHYS-PUB-2014-021}.
Two samples are generated with alternative CR models: in the `QCD-based' model (CR1) the formation of dipoles containing three quarks is enhanced and as a consequence the production of baryons is enhanced, while in the `gluon-move' model (CR2) only the gluons are considered for the reconnection.
To isolate the systematic effects related to the CR model, a third sample (CR0) is generated with the nominal MPI-based CR model, but employing a dedicated CR tune that is consistent with the tuning used in the CR1 and CR2 samples~\cite{TOPQ-2019-01,ATL-PHYS-PUB-2017-008}.

The effects of using different parton shower and hadronisation models are evaluated using a sample produced with
\POWHEG interfaced to \HERWIG[7.2.1]~\cite{Bahr:2008pv,Bellm:2015jjp,Bellm:2017jjp,Bellm:2019zci} using the default set of tuned parameters and the \MMHT[lo]~\cite{Harland-Lang:2014zoa} PDF set. The settings in \POWHEG~are the same as in the nominal sample.

Two further \ttbar\ samples are used to test the robustness of the fit model. One sample is simulated using the same setup as the nominal sample, except for having the early resonance decay parameter set such that resonance decays can occur before CR happens in the simulation. The other \ttbar\ sample is simulated with the \SHERPA[2.2.12]~\cite{Bothmann:2019yzt} generator using MEs at NLO in QCD for up to two partons, and at LO in QCD for up to four partons calculated with the \textsc{Comix}~\cite{Gleisberg:2008fv} and \OPENLOOPS~\cite{Buccioni:2019sur,Cascioli:2011va,Denner:2016kdg} libraries. They were matched with the \SHERPA parton shower~\cite{Schumann:2007mg} using the \MEPSatNLO prescription~\cite{Hoeche:2011fd,Hoeche:2012yf,Catani:2001cc,Hoeche:2009rj} and a dedicated set of tuned parton-shower parameters developed by the \SHERPA authors. The \NNPDF[3.0nnlo]~\cite{Ball:2014uwa} set of PDFs was used.

All \ttbar\ samples are normalised to the \mt-dependent cross-section prediction at next-to-next-to-leading-order (NNLO) in QCD, including a resummation of next-to-next-to-leading logarithmic (NNLL) soft-gluon terms.
\TOPpp[2.0] yields $\sigma_{t\bar{t}} = 832^{+20}_{-29}(\mathrm{scale})^{+35}_{-35}(\mathrm{PDF}+\alpha_{S})$~pb for $\mt=172.5$~\GeV~\cite{Beneke:2011mq,Cacciari:2011hy,Baernreuther:2012ws,Czakon:2012zr,Czakon:2012pz,Czakon:2013goa,Czakon:2011xx}.

Backgrounds to the signal process including the decay of at least one $W$~or $Z$~boson into leptons are simulated using MC samples. Top quark production in association with a $W$ boson ($tW$) and single-top-quark (single-top) production via the $s$-channel are modelled using the same setup as the nominal \ttbar~sample, using the five-flavour scheme. The diagram removal scheme~\cite{Frixione:2008yi} was used to remove interference and overlap between $tW$~and $\ttbar$\ production. An alternative sample is simulated using the diagram subtraction scheme~\cite{Frixione:2008yi,ATL-PHYS-PUB-2016-020} to evaluate the uncertainty due to the choice of the interference treatment scheme.
Another alternative sample is generated with \POWHEG interfaced to \HERWIG[7.04]~\cite{Bahr:2008pv,Bellm:2015jjp} to test the impact of a different parton shower and hadronisation model in $tW$\ production.
Single-top $t$-channel production is modelled in an identical way to $tW$ and $s$-channel production, but using the four-flavour scheme in the PDF set~\cite{Frederix:2012dh}. All single-top samples are normalised using cross-section predictions at higher orders, at NNLO in QCD for the $s$-channel and $t$-channel~\cite{Campbell:2020fhf} and at NLO+NNLL in QCD for the $tW$ process~\cite{Kidonakis:2021vob}.

The production of \ttV\ events, where $V$~represents a $W$ or $Z$~boson, is modelled at NLO in QCD using the
\MGNLO[2.3.3]~\cite{Alwall:2014hca} generator with the \NNPDF[3.0nlo] PDF set. The events were interfaced to \PYTHIA[8.210] using the A14 tune and the \NNPDF[2.3lo] PDF set. The production of $t\bar{t}H$ events is modelled using the same generator setup as the nominal \ttbar sample. The $t\bar{t}W$ sample was normalised to a cross-section calculated at NNLO in QCD, including NLO electroweak corrections~\cite{Buonocore:2023ljm}, whereas the $t\bar{t}Z$ and $t\bar{t}H$ samples were normalised to cross-sections calculated at NLO in both the strong and electroweak couplings~\cite{deFlorian:2016spz}.

The production of \(V+\)jets is simulated with the \SHERPA[2.2.1]~\cite{Bothmann:2019yzt} generator in an identical way to the \ttbar sample simulated using \SHERPA. The samples were normalised to a NNLO prediction in QCD~\cite{Anastasiou:2003ds}.

Samples of diboson final states (\(VV\)) are simulated with the \SHERPA[2.2.1] or 2.2.2~\cite{Bothmann:2019yzt} generator depending on the process, using MEs at NLO in QCD for up to one additional parton and at LO in QCD for up to three additional parton emissions. Samples for the loop-induced processes \(gg \to VV\) were simulated using LO-accurate ME for up to one additional parton emission. The ME calculations were matched and merged with the \SHERPA parton shower as in the \(V+\)jets samples. The samples also use the same set of PDFs and dedicated set of tuned-parameters.

All simulated samples other than those generated with \SHERPA use the \EVTGEN program~\cite{Lange:2001uf} to simulate the decays of bottom and charm hadrons. Most of the samples use version 1.6.0 of the program.
The alternative \ttbar samples used to probe the FSR and the parton shower and hadronisation uncertainties use version 1.7.0 and the $t\bar{t}W$ and $t\bar{t}Z$ samples are produced using version 1.2.0.

The QCD multijet background refers to events originating from processes that do not involve leptons produced by $W$, $Z$~or Higgs boson decays. It is estimated by using the data-driven `Matrix method'~\cite{EGAM-2019-01,TOPQ-2019-23}, in which a selection of events in data with looser identification and isolation lepton requirements is used along with measurements of the probability of leptons to satisfy the nominal selection requirements.


\section{Event reconstruction and selection}
\label{sec:eventSel}

Electron candidates are identified by matching energy deposits in the electromagnetic calorimeter with a corresponding track in the inner tracking detector. They must have $\pt>27$~\GeV and $|\eta|<2.47$, excluding candidates with $1.37 < |\eta| < 1.52$. They must also satisfy the `Tight' identification and isolation requirements discussed in Ref.~\cite{EGAM-2018-01}. Muon candidates are reconstructed by combining a track in the inner tracking detector with either a track or hits in the muon spectrometer. They are required to have $\pt>~27$~\GeV, $|\eta|<2.5$, satisfy the `Medium' identification requirements and the `Tight' isolation requirements~\cite{MUON-2018-03}. The transverse momenta requirements on the leptons are chosen to be slightly above the single-lepton trigger thresholds.
Corrections are applied to the simulation to bring the efficiency of lepton reconstruction and identification in agreement with the data and to correct the energy (momentum) scale and resolution of electrons (muons).

The \antikt clustering algorithm~\cite{Cacciari:2008gp,Fastjet} is used to reconstruct jets with radius parameter $R$~=~0.4 using particle flow objects, utilising both track and calorimeter information~\cite{PERF-2015-09}.
The energy scale of jets is calibrated using both simulation and data~\cite{JETM-2018-05} and selected jets must have $\pt>26$~\GeV and $|\eta|<2.5$.
Corrections are also applied to the simulation to bring the jet energy resolution in simulation into agreement with the jet energy resolution measured in data~\cite{JETM-2018-05}.
For jets with $\pt<60$~\GeV and $|\eta|<2.4$, a threshold placed on the output of the multivariate jet-vertex tagger (JVT) must be satisfied to remove jets originating from pileup~\cite{PERF-2014-03}.
Overlap removal criteria are applied to avoid double counting energy between leptons and jets, as discussed in Ref.~\cite{TOPQ-2019-23}.
Jets including $b$-hadrons ($b$-jets) are identified using the DL1r multivariate $b$-tagging algorithm~\cite{FTAG-2019-07}, where the chosen working point corresponds to an efficiency of 77\% per $b$-jet, and rejection factors of 170 and 5 for light-flavour ($u,d,s$) jets and jets initiated by $c$-quarks, respectively. Reconstructed jets satisfying this working point are referred to as $b$-tagged jets.
Measurements of the efficiency of the $b$-tagging algorithm are used to adjust the simulation to match the performance measured in data~\cite{FTAG-2018-01,FTAG-2020-08,FTAG-2019-02}.

To reconstruct the hadronic decay of boosted ($\pt \gtrsim 2\mt$) top quarks, the \antikt algorithm is used to form large-$R$~jets with radius parameter $R$~=~1.0 using as input selected $R$~=~0.4 (small-$R$) jets with $\pt>30$~\GeV~\cite{Nachman_2015}. A trimming procedure~\cite{TOPQ-2019-23,Nachman_2015,Krohn_2010} is applied to the large-$R$ jets to remove constituent jets likely to originate entirely from pileup.
The four-momenta of large-$R$~jets are calculated from the sum of four-momenta of the constituent jets, allowing the jet energy scale of the small-$R$ jets to be propagated to the large-$R$~jets.
To ensure that the large-$R$ jet corresponds to the hadronic decay of a boosted top quark it must have $\pt>355$~\GeV, $|\eta|<2.0$, contain at least two jet constituents where at least one must be a $b$-tagged jet, and have an invariant mass in the range of $120<\mtopjet<220$~\GeV. In cases where events have more than one reconstructed large-$R$ jet that satisfies these criteria, the one with highest \pt is associated with the boosted top-quark decay, and is referred to as the `top-jet'. The missing transverse momentum, with magnitude \met, is reconstructed based on the negative vector sum of the momentum of the calibrated leptons, small-$R$ jets with \pt~above $20$~\GeV ($30$~\GeV) for $|\eta|<2.5$~$(|\eta|>2.5)$~and the soft radiation term~\cite{JETM-2020-03}.

Selected events must have one top-jet ($J$), at least one $b$-tagged jet not associated with the top-jet, and exactly one lepton candidate. The selected lepton must match the lepton reconstructed by the trigger within $\Delta R<0.15$~\cite{TRIG-2018-01,TRIG-2018-05}.
A requirement of $\Delta R(e,J)>1.0$ ensures that the top-jet is not originating from a high-\pt electron.
Additionally, the lepton is required to be close to a $b$-tagged jet not associated with the top-jet, $\Delta R(\ell,b)<2.0$.
A requirement that the invariant mass of the lepton and the closest $b$-tagged jet satisfying the former selection must be less than $180$~\GeV is applied to reject events originating from $tW$ single-top production while retaining on-shell \ttbar signal events~\cite{TOPQ-2017-05}.
The contribution from the multijet background is reduced by requiring that events have $\met>20$~\GeV and $\met+m_{\text{T}}^W>60$~\GeV, where $m_{\text{T}}^W$ is the transverse mass of the $W$~boson.\footnote{$m_{\text{T}}^W~=~\sqrt{2p_{\text{T}}^{\ell}\met\left(1-\cos{\Delta\phi}\right)}$ where $p_{\text{T}}^{\ell}$ is the transverse momentum of the lepton and $\Delta\phi$~is the difference in azimuthal angle  between the lepton and missing transverse momentum.} These requirements ensure that the events have the expected boosted topology.


\section{Systematic uncertainties}
\label{sec:syst}

Many systematic uncertainties affect the measurement of \mt\ and are described below.
Systematic uncertainties are found to have approximately symmetric effects on the observables used in this Letter and are explicitly symmetrised as follows.
For uncertainties with separate `up' and `down' variations, the symmetric effect is obtained from the average of the two variations.
For systematic uncertainties where there is only one variation, this is taken as the `up' variation and the effect is mirrored to obtain the `down' variation.
Before measuring \mt, all systematic uncertainties are assumed to be uncorrelated with one another.
The impact of the uncertainties on the measurement
is discussed in Section~\ref{sec:fit}.

The uncertainty in the integrated luminosity is 0.83\%~\cite{DAPR-2021-01} and is applied to all processes other than the QCD multijet background.
The reweighting of the simulation to match the pileup distribution in data involves rescaling the average number of interactions per bunch crossing to achieve
improved agreement between data and simulation for the observed number of primary vertices. The uncertainty in this rescaling is propagated to the reweighting
factors to estimate the uncertainty due to pileup.

Electrons and muons have uncertainties associated with the efficiency for the reconstruction, identification and isolation requirements, all derived from studies using $Z\rightarrow ee$ and $Z\rightarrow\mu\mu$ events~\cite{EGAM-2018-01,MUON-2018-03}. Similar studies give the uncertainties in the trigger efficiencies~\cite{TRIG-2018-05,TRIG-2018-01}.
Uncertainties corresponding to the electron and muon energy/momentum scales and resolutions are evaluated using resonance decays~\cite{EGAM-2021-02,MUON-2022-01}.

Uncertainties in the jet energy scale~(JES) are evaluated using 35 independent variations in the jet energies that parametrise the uncertainties in the JES of $R=0.4$~particle flow jets~\cite{JETM-2018-05,STDM-2020-04,ATL-PHYS-PUB-2022-021}, which are
propagated through to the \largeR\ jets. Included are six variations related to the flavour of the jets.
Three variations relate to the response difference between light-flavour quark- and gluon-initiated jets and three for the response of heavy-flavour ($c,b$) quark-initiated
jets~\cite{STDM-2020-04,ATL-PHYS-PUB-2022-021}. The total flavour response uncertainty on $b$-quark-initiated jets is $1.2$\% ($0.3$\%) at $\pt = 30 (100)$~\GeV. This is an improved treatment relative to previous ATLAS \mt\ measurements and reduces the impact of flavour uncertainties in the measurement.
The remaining variations correspond to the uncertainties from the in situ calibration and from the impact of pileup on the jets.
Previous studies have shown that propagating the $R=0.4$~JES uncertainties to large-$R$ reclustered jets provides an accurate model of the large-$R$ jet energy and mass scales~\cite{ATLAS-CONF-2017-062,ATLAS-CONF-2020-022}.
In the remainder of the Letter, the heavy-flavour (HF) JES uncertainties are presented separately to the other JES uncertainties, as these are not expected to be constrained in the fit.
Uncertainties in the jet energy resolution (JER) are evaluated using 13 independent variations that reflect the uncertainties in the measurements of the JER~\cite{JETM-2018-05} and these
are propagated to the \largeR\ jets.
An uncertainty in the efficiency of the JVT requirement on jets for pileup suppression is also included.

Uncertainties corresponding to the performance calibrations of the $b$-tagging algorithm in data are propagated by varying the efficiency corrections associated with the $b$-tagging~\cite{FTAG-2018-01,FTAG-2020-08,FTAG-2019-02}. This uncertainty model includes nine / four / four independent variations for the $b$\ / $c$\ / light-jet calibrations, respectively, and two components for the extrapolation to high-\pt jets.

Uncertainties in the \met arise from potential mismodelling of the tracks in the soft term. Three variations for the uncertainty in the scale and resolution of the \met\ soft term are included.
Furthermore, the uncertainties in the energy scales and resolutions of the leptons and jets are propagated to the \met.

The modelling of \ttbar\ events is a complex process and several systematic uncertainties are included to reflect the corresponding uncertainties.
These are assessed by either reweighting the nominal \ttbar~sample or by using the alternative \ttbar~samples described in Section~\ref{sec:dataMC}.
The uncertainty due to missing higher-order corrections is covered by separately varying \muR and \muF in the ME calculation by factors of 0.5 and 2.0.
The \pt\ distribution of the top quarks is known to be softer in data than the nominal \ttbar~prediction and better described by NNLO calculations~\cite{TOPQ-2018-15,CMS-TOP-20-001,TOPQ-2019-23}. As a consequence, an additional uncertainty is included by reweighting the distribution of the top quark \pt, and the \pt and mass distributions of the \ttbar pair in the simulation to match the prediction calculated at NNLO in QCD with NLO electroweak corrections~\cite{Czakon:2017wor}.
The uncertainty originating from the PDFs is evaluated using the 30 eigenvectors of the PDF4LHC PDF set~\cite{Butterworth:2015oua}.
Several systematic variations are made to cover uncertainties in the amount of QCD radiation in \ttbar\ events.
One uncertainty component is obtained by reweigthting the \ttbar\ sample
to use the Var3c eigentune of the \Pythia[8] A14 tune, which changes the scales in the initial-state radiation (ISR) shower in \PYTHIA.
Another uncertainty component is derived from
the \ttbar\ sample with the alternative value for the \hdamp\ parameter.
Variations in the FSR are obtained using the samples with alternative values of $\alpha_s^{\mathrm{FSR}}$.
The change in scale is applied to all branchings in the parton shower, consistent with the expectation that at high \pt light-flavour quarks and $b$-quarks must behave in the same way.
The uncertainty related to the matching of the NLO calculation is evaluated using the  \ttbar\ sample with the alternative $\pt^{\text{hard}}$ setting.
Uncertainties in the parton shower and hadronisation model are assessed by comparing the sample generated with \POWHER{7.2.1} with the nominal \ttbar\ sample.
In the simulation of QCD radiation from the $b$-quarks produced in the top quark decay, there is an ambiguity in the choice of the recoil particle for gluon emissions after the first emission~\cite{Brooks:2019xso}. This can affect out-of-cone radiation for the reconstructed $b$-tagged jets and hence the reconstructed top quark mass.
An uncertainty for this is assessed by using the sample with the alternative recoil setting in \Pythia.
The uncertainty from the underlying event is assessed using the two samples generated with the Var1 eigentune variation.
Uncertainties in the CR model are assessed by comparing the samples with the two alternative CR models (CR1 and CR2) with the CR0 sample.
CR1 and CR2 are seen to have different impacts on the observables used in the measurement. Both models have different $\mWreco$ distributions to the nominal, while only CR2 is seen to have a different shape in \mR (see Section~\ref{sec:obs} for definitions of \mR and $\mWreco$). Both comparisons are therefore included as systematic uncertainties.

The event selection (Section~\ref{sec:eventSel}) results in a sample dominated by \ttbar\ events and hence uncertainties in the background model play a minor role in the measurement.
The largest background is the single-top $tW$\ process. Variations of the scales in the NLO calculation and the parton shower are included in the same way as done for the \ttbar sample.
These uncertainties are assumed to be uncorrelated with the uncertainties in the \ttbar\ sample.
The uncertainty originating from the removal of \ttbar\ events from the $tW$~sample is accounted for by comparing a sample with the diagram subtraction scheme~\cite{Frixione:2008yi} with the nominal sample that uses the diagram removal scheme.
The uncertainty in the overall production rate of $3.7\%$ for $tW$\ includes scale, PDF and $\alpha_s$\ uncertainties in the higher-order calculation of the cross-section.
The equivalent uncertainties for the single-top $s$- and $t$-channel processes are $3.9\%$~and $1.9\%$, respectively.
The uncertainty in the $W+$jets background is evaluated by modifying event weights corresponding to varying the renormalisation and factorisation scales simultaneously
by factors of 0.5 and 2.0.
The background from processes where top-quark-top-antiquark pairs are produced in association with either a vector boson or the Higgs boson ($\ttbar X$) is dominated by the $\ttbar W$~process.
As the $\ttbar W$~cross-section recently measured by ATLAS~\cite{TOPQ-2019-30} is higher than the theoretical prediction by slightly more than the theory uncertainty, the difference between the
measurement and the theory ($18\%$) is used as the normalisation uncertainty for all the $\ttbar X$~processes.
Very few $Z$+jets and diboson events satisfy the event selection and an uncertainty of $50\%$~is used for the normalisation of these processes.
The QCD multijet background has a normalisation uncertainty of 65\%, following Ref.~\cite{TOPQ-2019-23}.


\section{Definition of observables}
\label{sec:obs}

Three observables are built using the kinematic information of the selected events. The first observable is the invariant mass of the top-jet, $\mtopjet$, whose mean value ($\meanmtopjet$), calculated using events with $\mbox{145 < \mtopjet < 205}$~\GeV, is sensitive to \mt, as shown in Figure~\ref{fig:observables:mJparam}.

The second observable is $\mWreco$, defined as the invariant mass of the two non $b$-tagged constituent jets inside the top-jet that have the largest $\pt$.
This observable is defined only for events where the constituents of the top-jet include at least two jets that are not $b$-tagged.
In the nominal $\ttbar$~simulation, $62\%$~of the selected events are expected to meet this requirement.
This observable aims at identifying the jets that originate from the $W\to q\bar{q}^\prime$~decay and is sensitive to the JES\@.
This is demonstrated in Figure~\ref{fig:observables:mjj}, which shows
the expected distribution for the observable and for the first eigenvector component of the JES systematic uncertainties relating to the MC models used for the in situ JES calibration~\cite{JETM-2018-05}. The expected distribution includes a clear peak corresponding to the $W$~boson decay, where the $W$~boson mass is sufficiently well known such that $\mWreco$ can be used to constrain the JES uncertainties.

The final observable is $\mR$, defined as the invariant mass of the semi-leptonically decaying top quark ($t\to \ell\nu b$) and the closest additional jet.
This observable is designed to select events where there is additional QCD radiation from the $b$-quark
and aims at constraining the systematic associated with the recoil in the top quark decay (see Section~\ref{sec:syst}).
The increased wide-angle radiation in the alternative recoil model results in lower $\meanmtopjet$~than the nominal model and this change in radiation pattern
also results in an increase in the number of events with an additional jet close to the semi-leptonically decaying top quark, allowing \mR to control this uncertainty source.
The semi-leptonically decaying top quark is reconstructed using the lepton, the closest $b$-tagged jet not associated with the top-jet, and the neutrino, where the missing transverse momentum is used to estimate $p^{\nu}_{x}$ and $p^{\nu}_{y}$, while $p^{\nu}_{z}$ is calculated using $m(\ell\nu)=\mW$~\cite{TOPQ-2019-23}.
Using the subset of events ($3.7\%$~of selected simulated $\ttbar$~events) that have an additional jet that is close to this reconstructed top quark, $\Delta R(t,j)<0.5$,
$\mR$ is expected to peak
close to the top quark mass for events with additional radiation from the $b$-quark.
The distribution of $\mR$~for the nominal simulation is compared with the one from the alternative recoil model in Figure~\ref{fig:observables:mtj}.
The alternative model is found to predict more events than the nominal model, particularly near $\mR\sim\mt$, consistent with the expectation of increased wide-angle radiation.

The numbers of events satisfying the selection criteria for the data and the simulated signal and background processes
for the three observables are shown in Table~\ref{table:prefitYields}.
The observed difference between the number of data events and the total prediction is consistent with the previous observations that the prediction from \POWPY is higher than the measured cross-section at high top quark \pt~\cite{TOPQ-2018-15,TOPQ-2019-23}.
The agreement between prediction and data improves if \POWPY~is reweighted to match NNLO predictions~\cite{TOPQ-2018-15}.

\begin{figure}[!htb]
\centering
\subfloat[]{\includegraphics[width=0.33\textwidth]{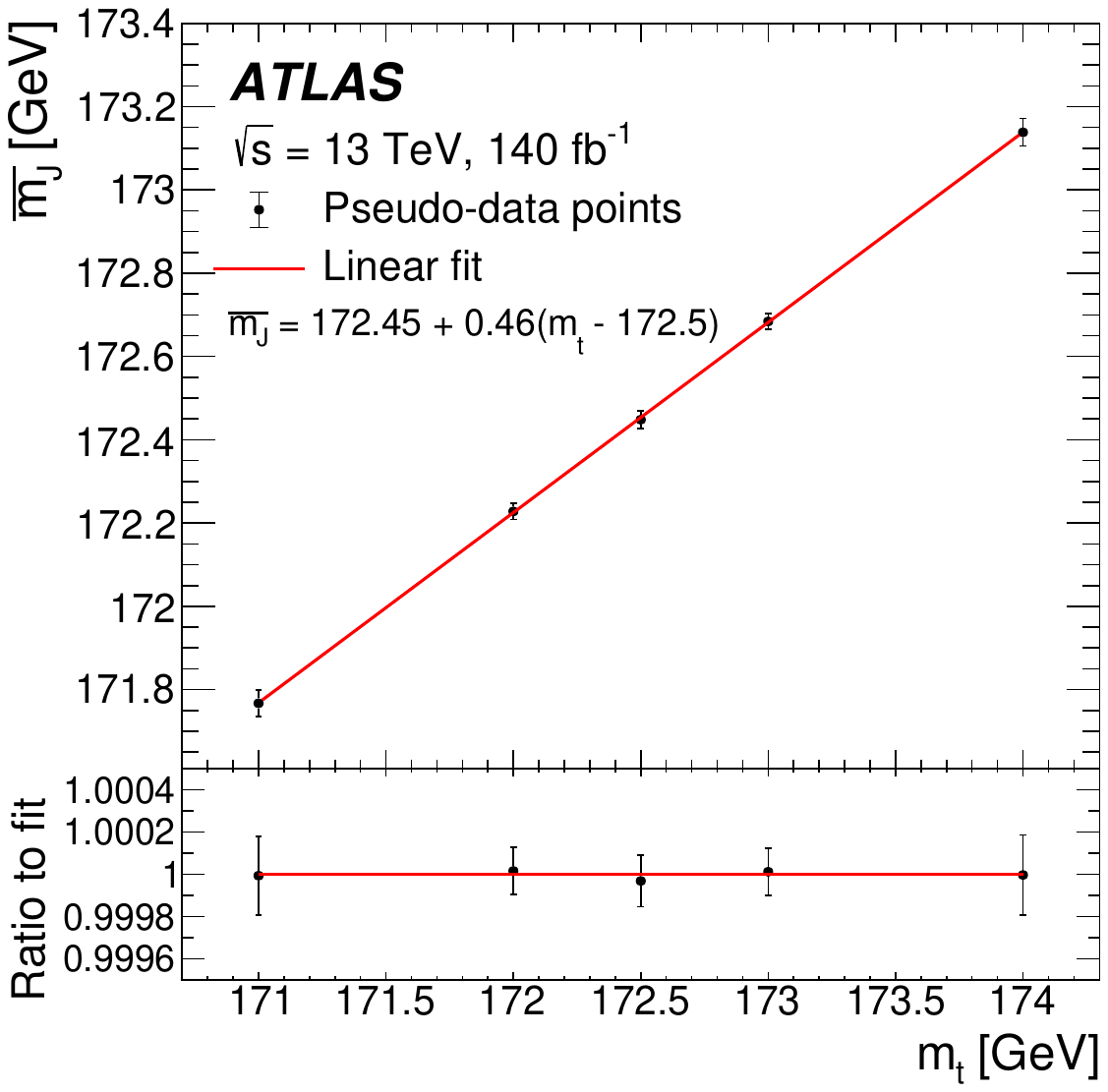}\label{fig:observables:mJparam}}
\subfloat[]{\includegraphics[width=0.33\textwidth]{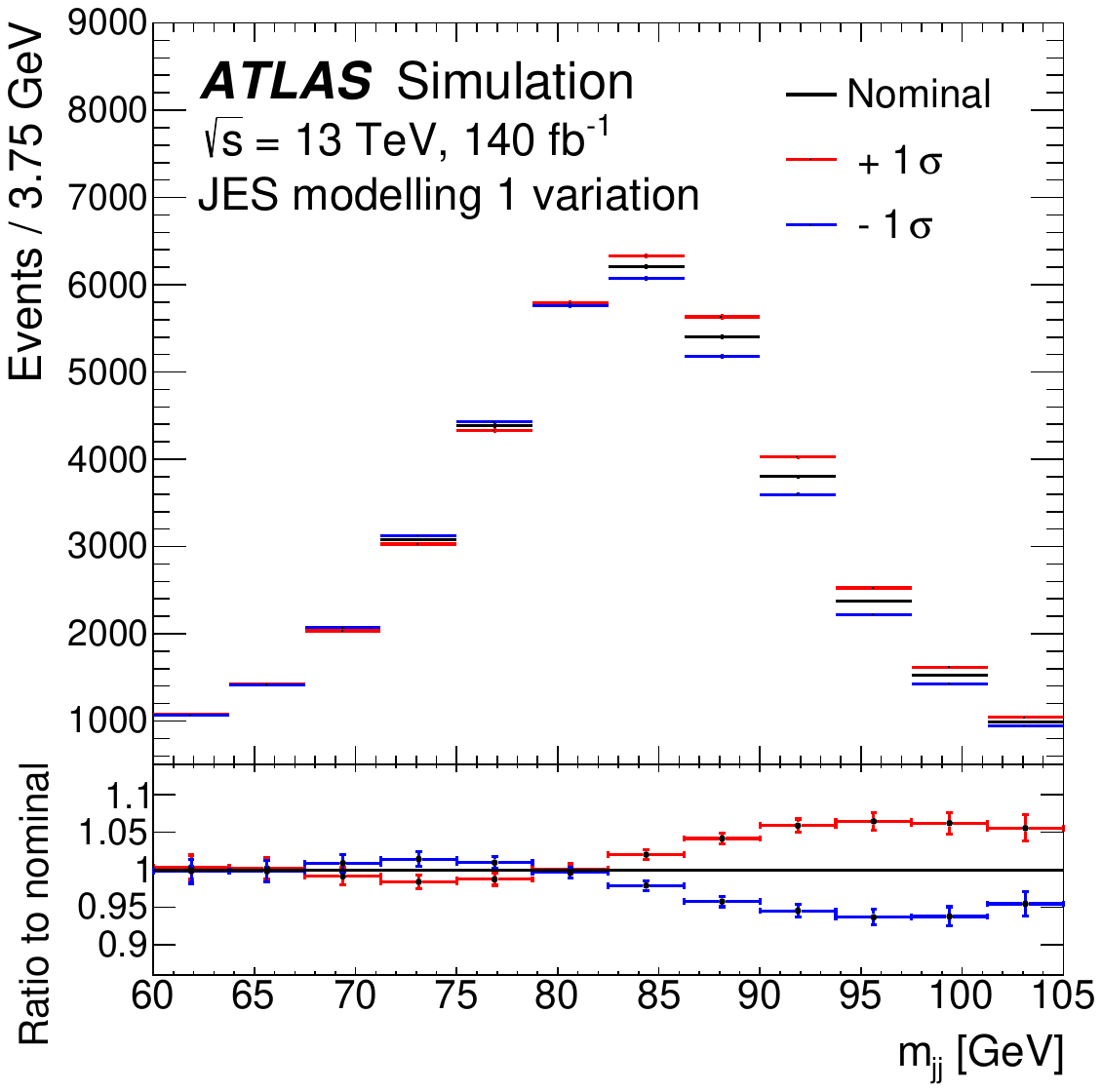}\label{fig:observables:mjj}}
\subfloat[]{\includegraphics[width=0.33\textwidth]{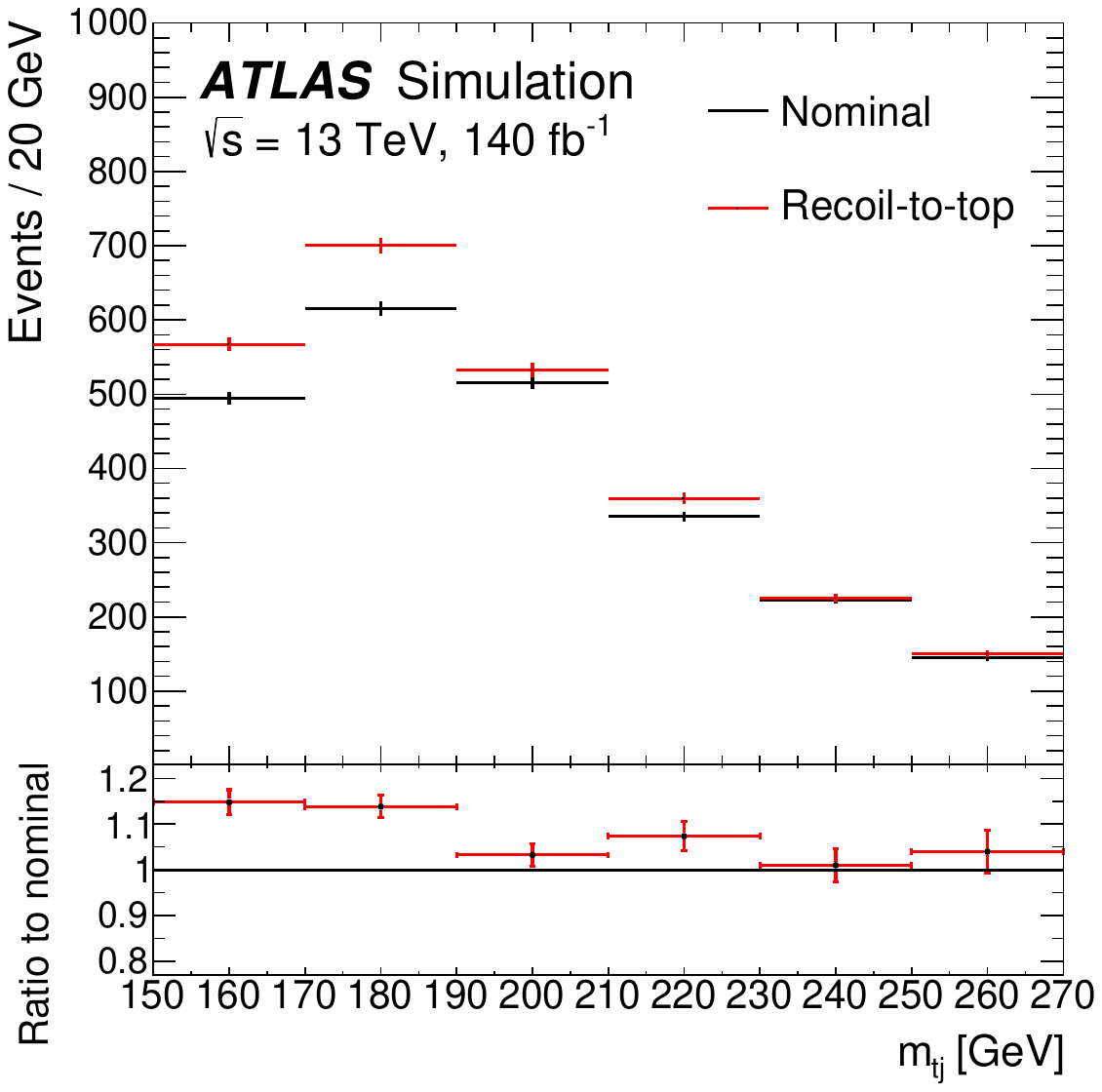}\label{fig:observables:mtj}}
\caption{The dependence of \meanmtopjet on \mt is shown in (a) where the filled pseudo-data points are built from \ttbar\ samples with different values of the top quark mass and the nominal background prediction. For each point, \meanmtopjet is calculated using events in the range of $145 < \mtopjet < 205$~\GeV. The linear fit to the points is shown by the solid line and the fit parameters are also displayed in (a). (b) Comparison between the $\mWreco$ distribution
expected for the nominal $\ttbar$~simulation and the simulation where one of the JES uncertainties is varied to  $\pm 1\sigma$~and
(c) \mR\ distribution for the alternative recoil-to-top model used in the top quark decay compared with the nominal \ttbar\ prediction.
The error bars show the statistical uncertainty in the MC samples. }
\end{figure}

\tableVSpace{1.3}
\begin{table}[!htb]
\caption{The observed number of data events is compared with the expectation from the signal and background processes for the selections for the \mtopjet, \mWreco and \mR observables. The uncertainties in the expectations include all systematic uncertainties described in Section~\ref{sec:syst}.}\label{table:prefitYields}
\centering
\begin{tabular}{lccc}\toprule
Process & $145 < \mtopjet < 205$~\GeV & $60 < \mWreco < 105$~\GeV  & $150 < \mR <270$~\GeV \\
\midrule
\ttbar & 65000~$\pm$~9500 & 38100~$\pm$~5800 & 2340~$\pm$~430\\
Single-top & 1000~$\pm$~170 & 400~$\pm$~130 & 50~$\pm$~9\\
$\ttbar X$ & 700~$\pm$~130 & 360~$\pm$~67 & 54~$\pm$~10\\
Multijet & 400~$\pm$~260 & 170~$\pm$~110 & 23~$\pm$~15\\
$W+$\,jets  & 250~$\pm$~100 & 59~$\pm$~22 & 21~$\pm$~9\\
$Z\,+$\,jets & 49~$\pm$~24 & 10~$\pm$~5 & 5~$\pm$~3\\
Diboson & 22~$\pm$~11 & 5~$\pm$~3 & 2~$\pm$~1\\
\midrule
Total prediction & 67400~$\pm$~9500 & 39100~$\pm$~5800 & 2500~$\pm$~430\\
\midrule
Data & 57459 & 32722 & 2312\\
\bottomrule
\end{tabular}
\end{table}



\section{Profile-likelihood fit and results}
\label{sec:fit}

The mean of the invariant mass of the selected top-jet is found to depend linearly on \mt as
$\meanmtopjet(\mt)~=~A+B\cdot(\mt-\text{172.5})$. The parameters $A$~and $B$~are extracted by performing a fit to combined signal and background templates
created from the \ttbar\ MC samples with different \mt values and the fit is displayed in Figure~\ref{fig:observables:mJparam}.
The effect of changing \mt in the single-top background was found to be negligible and is not included in the parameterisation.
The \meanmtopjet also depends on the number of $\ttbar$ events (due to the background contribution) and the systematic uncertainties, included via the nuisance parameters (NP)~$\boldsymbol{\theta}$.
Including these effects results in
\begin{equation}
\meanmtopjet(\mt,\muttbar,\boldsymbol{\theta})~=~A+B\cdot(\mt-\text{172.5}) + C\cdot(\muttbar - 1) + \sum_s\theta_s\Delta_s,
\end{equation}
where $\muttbar$ controls the expected number of \ttbar\ events and is defined such that $\muttbar=1$\ is the SM prediction with $\sigma_{\ttbar}=832$~pb (see Section~\ref{sec:dataMC}). The parameter $C=0.02$~\GeV is determined by varying the $\ttbar$\ cross-section by $\pm10\%$. This small impact on the top quark mass originates from the low number of background events. The parameters $\Delta_s$\ encode the symmetric impact of changing each systematic uncertainty $s$ by one standard deviation.
The simulation confirms that all systematic uncertainties with a significant impact on $\meanmtopjet$~have a symmetric impact on $\meanmtopjet$.
Due to the large number of selected events, the measured value in data $\meanmtopjet^d$\ is expected to follow a Gaussian distribution ($G$) with width $\sigma_{\meanmtopjet}$\ given by the root-mean-square (RMS) of \meanmtopjet divided by the square root of the expected number of events.
A likelihood is constructed using \meanmtopjet and the binned
distributions of \mWreco and \mR:
\begin{align}
\label{eq:Final3DLikelihoodEquation}
L\left(\meanmtopjet^{d},\boldsymbol{{n}_{\mWreco}},\boldsymbol{{n}_{\mR}}\vert \mt,\muttbar,\boldsymbol{\mathbf{\theta}}\right)= & ~G\left[\meanmtopjet^{d}\vert\meanmtopjet(\mt,\muttbar,\boldsymbol{\mathbf{\theta}}),\sigma_{\meanmtopjet}\right]  \nonumber \\
& \times \prod_i P\left[n_{\mWreco,i}\vert \nu_{i}(\muttbar, \boldsymbol{\theta})\right]
\times \prod_k P\left[n_{\mR,k}\vert \rho_{k}(\muttbar, \boldsymbol{\theta})\right] \\
& \times \prod_s G\left[\beta_s\vert\theta_s,1\right], \nonumber
\end{align}
where $\nu_{i}$ and $\rho_k$ are the expected number of events in each bin of the \mWreco\ and \mR\ observables, respectively.
The number of bins is chosen to achieve sufficient sensitivity to the systematic uncertainties, while retaining enough events in each bin to avoid issues with statistical uncertainties in the simulation.
Twelve bins
are used for \mWreco\ and three for \mR.
The large bin width for \mR\ ensures negligible dependence on \mt~and \mWreco\ also has no visible dependence on \mt.
Consequently, $\nu_{i}$ and $\rho_k$~depend on the number of $\ttbar$\ events and the systematic uncertainties, but not \mt.
The observed number of events in the bins, $\boldsymbol{{n}_{\mWreco}}$~and $\boldsymbol{{n}_{\mR}}$~act to constrain the impact of the systematic uncertainties on the \mt measurement.
The events contributing to \meanmtopjet, \mWreco and \mR\ are not orthogonal and there is a degree of statistical correlation between the observables.
The likelihood assumes that the observables can be treated as statistically uncorrelated such that $\nu_{i}$ and $\rho_k$ follow Poisson distributions ($P$).
This assumption is validated with the test discussed in the next paragraph.
A pruning procedure is used to remove systematic uncertainties that have a negligible impact on the measurement. Any systematic uncertainties that change
the normalisation of both \mWreco~and \mR~by less than $0.1\%$~and cause the event yield in every bin of \mWreco~and \mR~to change by less than $0.01\%$~(not including normalisation effects)~are not included in the likelihood fit.
The effect of the systematic uncertainties on the \mWreco\ distribution is assumed to be smooth and hence a smoothing procedure is employed to remove statistical fluctuations from the systematically varied templates.
The small number of bins means that no smoothing is needed for \mR.
The final term of Eq.~(\ref{eq:Final3DLikelihoodEquation}) contains the prior constraints on the systematic uncertainties, defined such that the $\beta_s$\ parameters are zero and the corresponding Gaussian distributions have width equal to one.
A Poisson term is included into the likelihood for each bin in the \mWreco\ and \mR\ distributions to account for statistical uncertainties in the MC simulation; these terms are not displayed in Eq.~(\ref{eq:Final3DLikelihoodEquation}).
The top quark mass and its uncertainty are finally extracted from the profile-likelihood ratio~\cite{Cowan:2010js}:
\begin{equation}
\lambda(\mt) = \frac{ L\left(\meanmtopjet^{d},\boldsymbol{{n}_{\mWreco}},\boldsymbol{{n}_{\mR}}\vert \mt,\hathat{\muttbar},\hathat{\boldsymbol{\mathbf{\theta}}}\right) }{ L\left(\meanmtopjet^{d},\boldsymbol{{n}_{\mWreco}},\boldsymbol{{n}_{\mR}}\vert \hat{\mt},\hat{\muttbar},\hat{\boldsymbol{\mathbf{\theta}}}\right) },
\end{equation}
where $\hat{\mt},\hat{\muttbar},\hat{\boldsymbol{\mathbf{\theta}}}$\ are the parameters that maximise the likelihood and $\hathat{\muttbar},\hathat{\boldsymbol{\mathbf{\theta}}}$\ maximise the likelihood for the given value of \mt.

The performance of the likelihood fit is evaluated before examining the data.
The assumption that the observables can be treated as uncorrelated
is tested by performing pseudo-experiments that reflect the statistical uncertainty and include the statistical correlation between the observables.
Each pseudo-experiment is fitted in turn and the RMS of the fitted \mt values is found to agree with the results of pseudo-experiments that ignore the correlations. The mean of the pseudo-experiments is also found to agree with the input \mt.
The linearity is tested by injecting pseudo-data into the fit corresponding to the expectation from using combined signal and background samples created from the \ttbar\ MC samples with different \mt values.
This test probes whether the \mWreco and \mR distributions can be treated as independent from \mt and the validity of neglecting the dependence of the background on \mt.
The fit linearity is found to be excellent and a residual fit closure uncertainty of $0.07$\ \GeV\ is assigned to cover deviations from perfect linearity.

The inclusion of \mWreco~and  \mR\ in the fit improves the total expected uncertainty
from $\totalUncAsimovMJ$~\GeV\ to $\totalUncAsimov$~\GeV\ for $\mt=172.5$~\GeV.
The largest improvement comes from the reduction of the total JES uncertainty from $1.4$~\GeV~to $0.26$~\GeV.
The inclusion of the \mWreco~distribution also reduces the impact of the ISR and FSR uncertainties on the measurement,
where the expected uncertainty contribution reduces from $0.82$~\GeV\ in a fit with only \meanmtopjet\ to $0.14$~\GeV in the full fit.
The inclusion of the \mR\ distribution reduces the recoil systematic uncertainty
from $0.36$~\GeV\ in a fit with \meanmtopjet~and \mWreco\ to $0.08$~\GeV\ in the full fit.
The inclusion of \mWreco and \mR increases the statistical uncertainty in $\mt$~from
$0.11$~\GeV to $0.25$~\GeV, but this increase is more than compensated by the reduction of the total systematic uncertainty in $\mt$~from $1.7$~\GeV~to $0.44$~\GeV.
The robustness of the fit model to $\ttbar$~modelling uncertainties is tested by fitting pseudo-data created from the nominal background expectation and either the \Sherpa~\ttbar sample or
the \POWPY~\ttbar~sample with the alternative early resonance decay setting (see Section~\ref{sec:dataMC}).
In each case, the fit returns \mt~with a bias smaller than $0.08$~\GeV, which is significantly smaller than the total expected modelling uncertainty in the measurement ($0.27$~\GeV).
This demonstrates the fit is able to adjust to $\ttbar$~MC models that are not directly included in the fit model as systematic uncertainties.

\begin{figure}[!htb]
\centering
\subfloat[]{\includegraphics[width=0.34\textwidth]{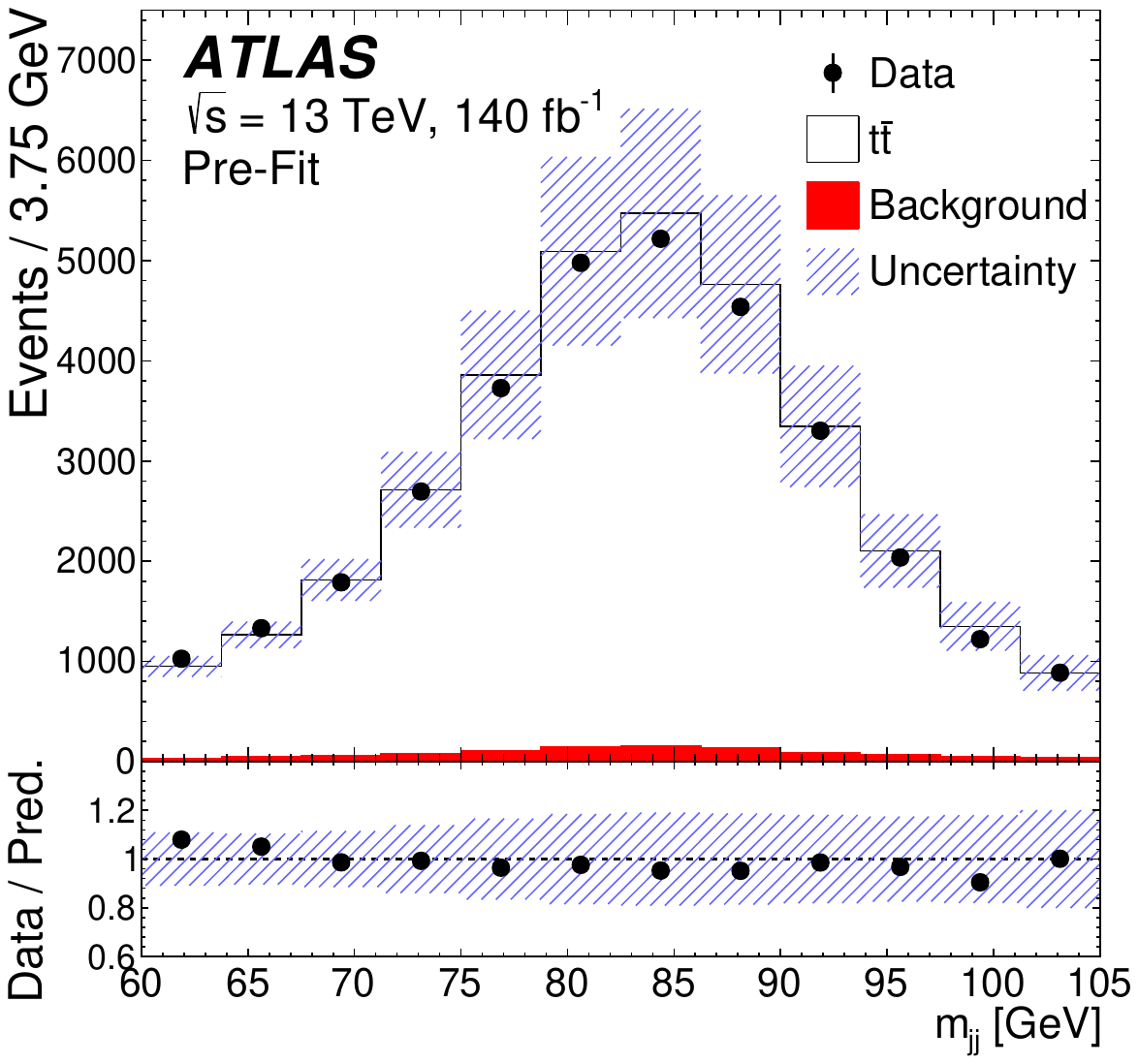}\label{fig:mjjprefit}}
\subfloat[]{\includegraphics[width=0.34\textwidth]{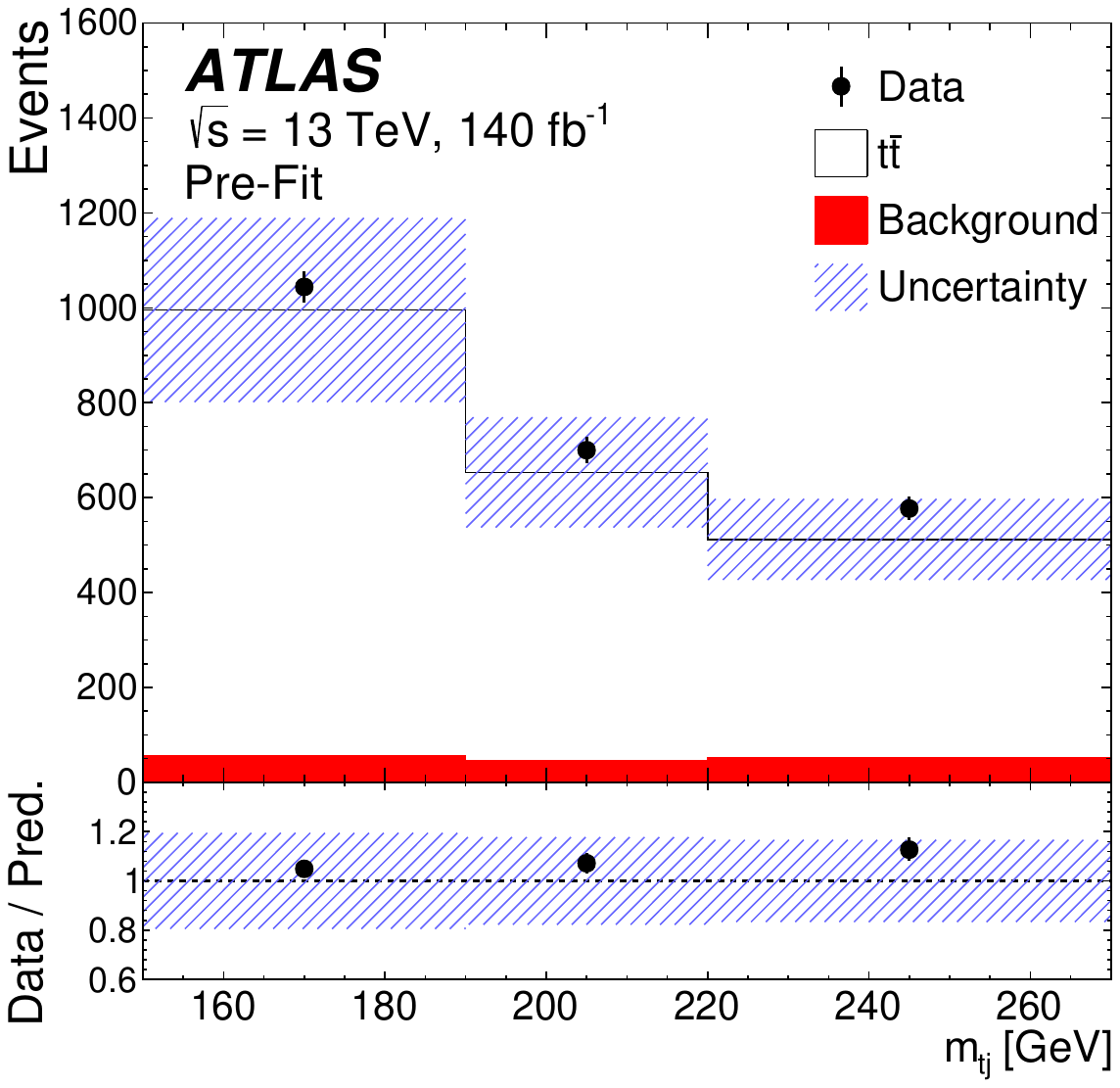}\label{fig:mtjprefit}}\\
\subfloat[]{\includegraphics[width=0.34\textwidth]{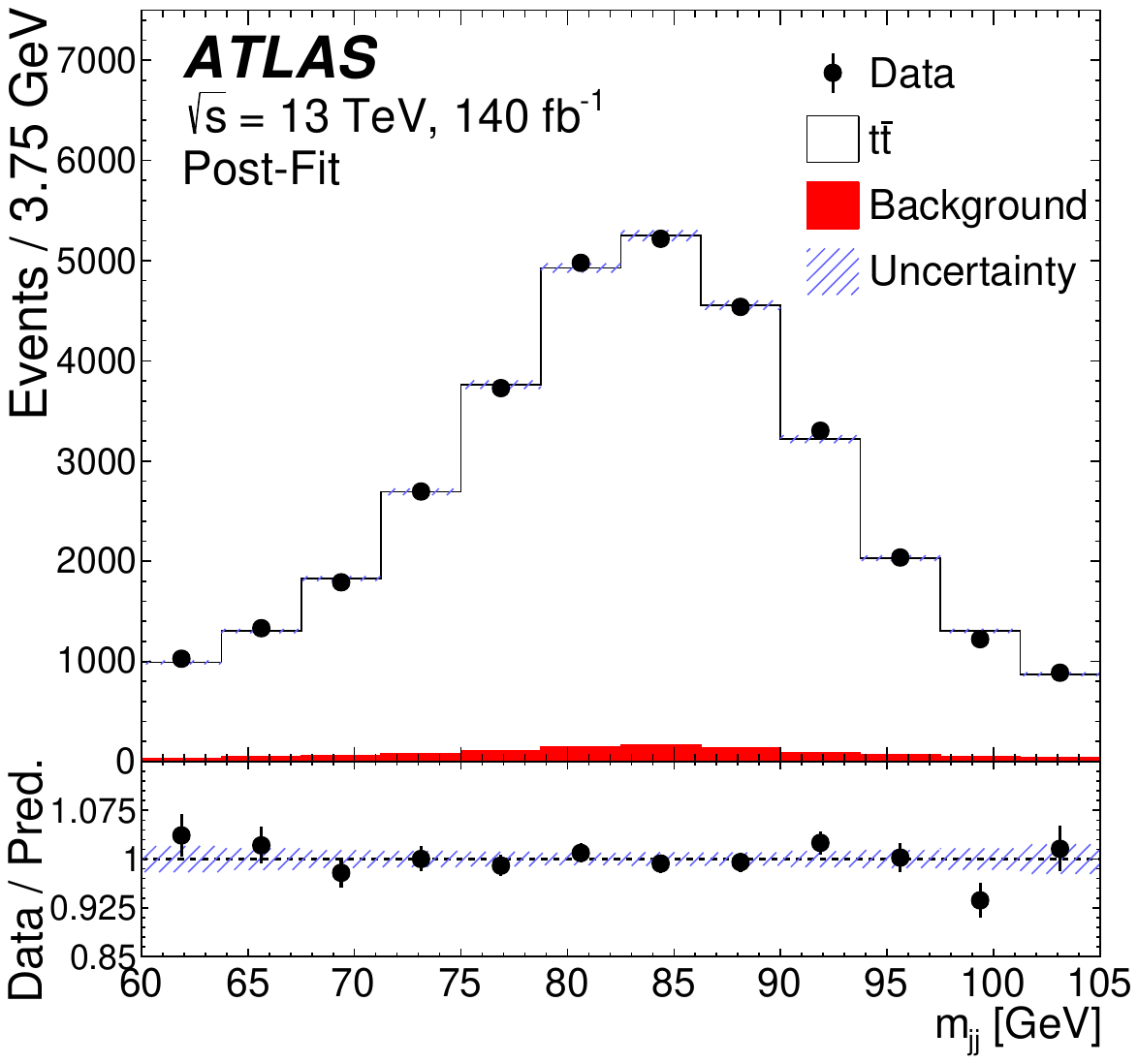}\label{fig:mjjpostfit}}
\subfloat[]{\includegraphics[width=0.34\textwidth]{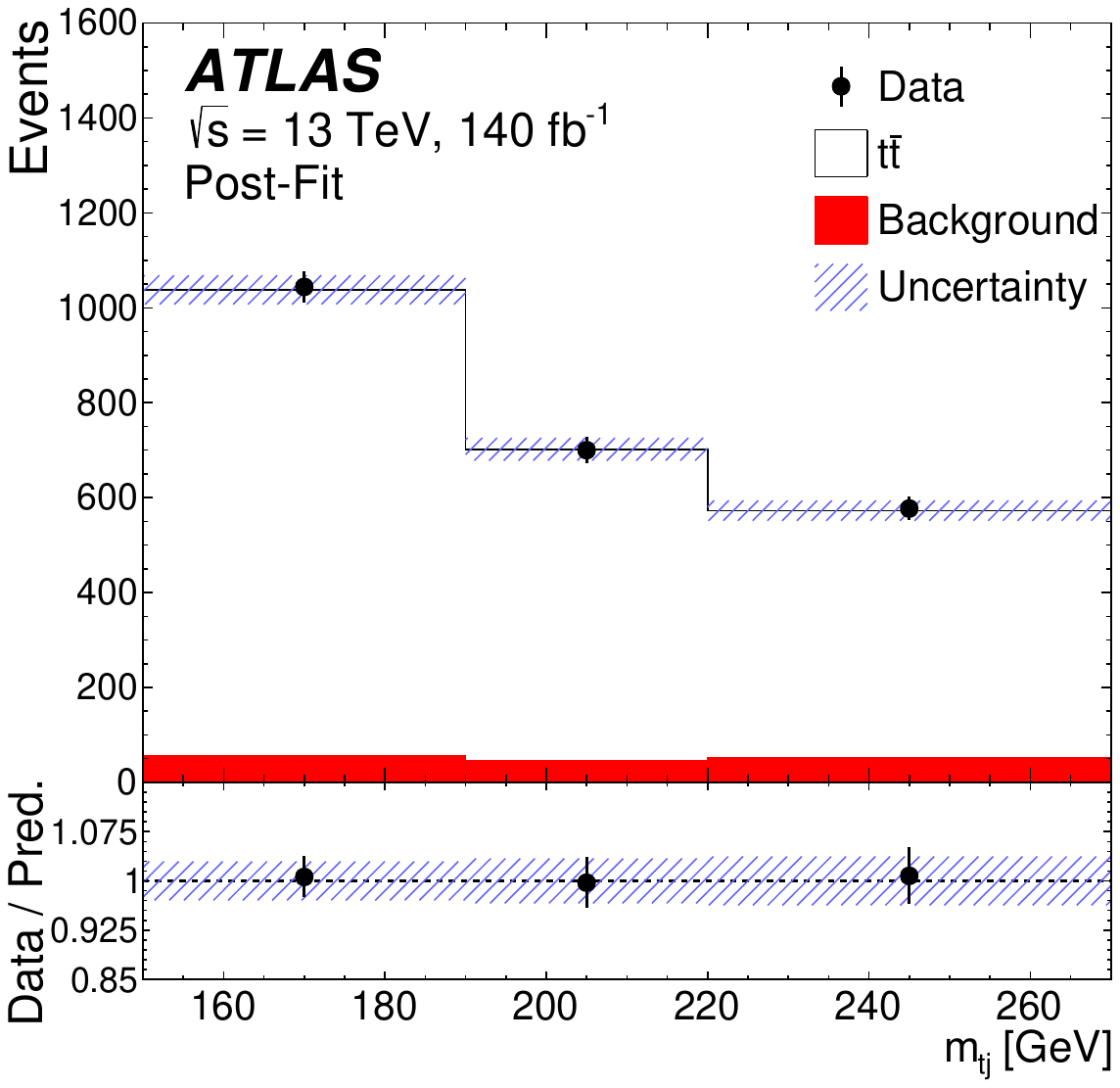}\label{fig:mtjpostfit}}
\caption{The pre and post-fit expectations for the \mWreco and \mR~distributions are compared with the data.
For the pre-fit case, the $\ttbar$\ sample corresponds to $\mt=172.5$~\GeV and is scaled to $\muttbar=0.856$, which is the ratio of data to theory found in the previous ATLAS cross-section measurement~\cite{TOPQ-2019-23}.
There is no uncertainty in the pre-fit value of \muttbar.
The uncertainty band represents the total uncertainty in the
model and is significantly reduced after the fit.
The bottom panel reflects the agreement between the data and total prediction, where the error bars show the statistical uncertainty of the data. This includes the systematic and MC statistical uncertainties and for the post-fit case it also includes the uncertainty on  \muttbar.
}
\end{figure}

The use of the \meanmtopjet, rather than the full distribution of \mtopjet means that the pulls and constraints on the NP could be examined by performing
a fit to data using only $\mWreco$\ and \mR.
The full fit adds one additional observable and one free parameter, meaning the fitted values of the NPs do not change when \meanmtopjet is added to the fit
of $\mWreco$\ and \mR.
The value of \meanmtopjet in data was blinded with a random offset until the full fit model was finalised.
The pre-fit distributions for \mWreco and \mR are compared with the data in Figures~\ref{fig:mjjprefit} and~\ref{fig:mtjprefit} and the corresponding post-fit distributions are shown in Figures~\ref{fig:mjjpostfit} and~\ref{fig:mtjpostfit}.
The \ttbar\ normalisation factor is found to be $\mu=0.87 \pm 0.09$, in good agreement with the results of the dedicated cross-section measurement~\cite{TOPQ-2019-23}.
The model is seen to provide a good description of the data, with a goodness-of-fit probability of $0.36$~obtained using the saturated model~\cite{SaturatedModel} and the fit improves the agreement in the shape of the observables.
The fitted values for the NPs with the largest contributions to the total uncertainty \mt are displayed in Figure~\ref{fig:np}.
No pulls beyond one standard deviation are observed, while some NPs are constrained, indicating the fit provides valuable information about the systematic uncertainties.
The parton shower and hadronisation systematic uncertainty has a large effect on the \mWreco distribution, which results in a significant constraint on the corresponding NP\@.
As this uncertainty encompasses more than one physics effect, propagating the constraint from \mWreco to \mtopjet and \mR is avoided by decorrelating this systematic uncertainty between the observables. The post-fit values for this systematic uncertainty for the three observables all agree with each other,
indicating that the nominal model based on \Pythia agrees consistently with data across all three observables.
The central value of \mt\ is stable with respect to this choice, changing from the nominal result by $0.06$~\GeV if the fit is performed with a single correlated parton shower and hadronisation NP\@.
The recoil NP is constrained due to the inclusion of the \mR distribution. The fitted value is found to be in excellent agreement with the model in the nominal \Pythia sample and
the uncertainty in the NP is significantly reduced in the fit. Moderate pulls and constraints are obtained on several NPs associated with JES uncertainties, which is expected as \mWreco~is
directly sensitive to the JES\@.
The NP for FSR is also constrained, which is expected as this physics effect causes changes in the shape of the \mWreco distribution.

\begin{figure}[!htb]
\centering
\hspace*{-2cm}
\includegraphics[width=0.6\textwidth]{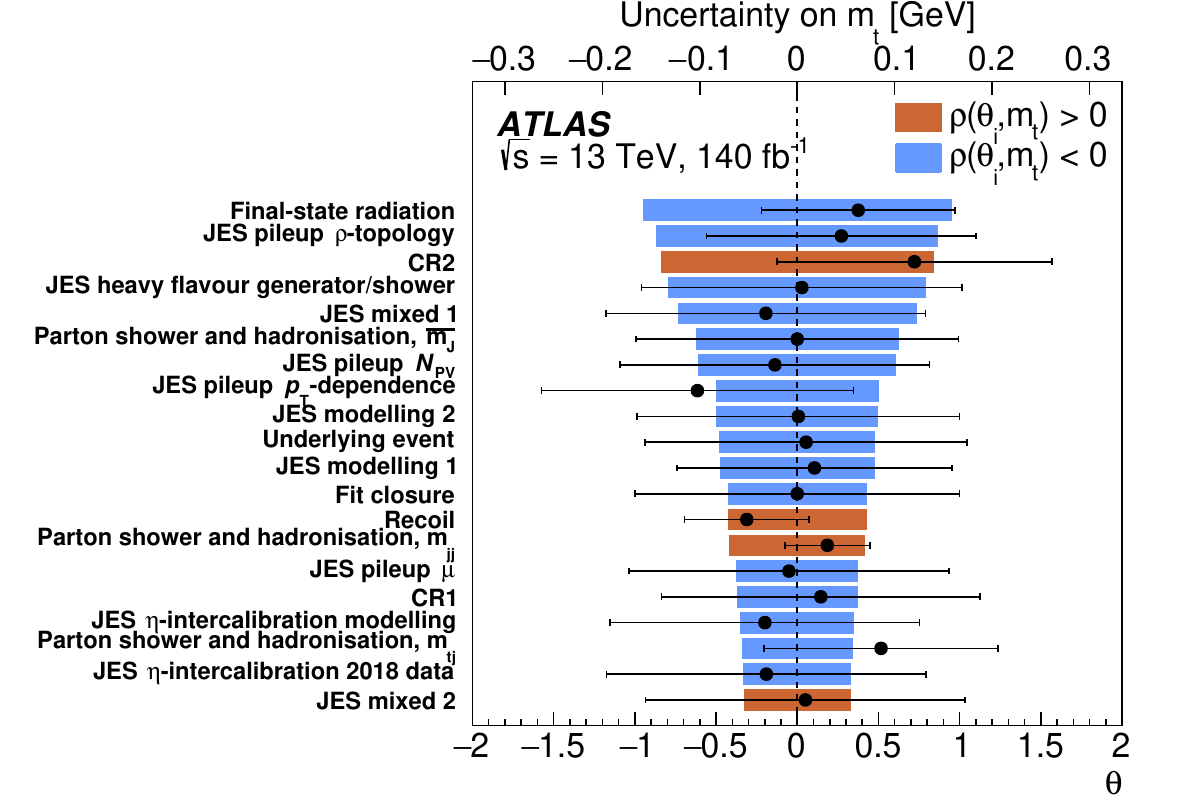}
\caption{
The contribution of individual NPs, $\theta$, to the total uncertainty in \mt is shown by the coloured bars and can be read off using the upper $x$-axis.
The red (blue) coloured bars represent NPs that have a positive (negative) correlation with \mt~and only NPs that have a contribution to the uncertainty in \mt of more than $0.05$~\GeV\ are shown.
The fitted values and post-fit uncertainties for these NPs are shown by the points, which refer to the lower $x$-axis.
All modelling uncertainties displayed in the figure correspond to uncertainties in the \ttbar\ process.}
\label{fig:np}
\end{figure}

\begin{figure}[!htb]
\centering
\subfloat[]{\includegraphics[width=0.32\textwidth]{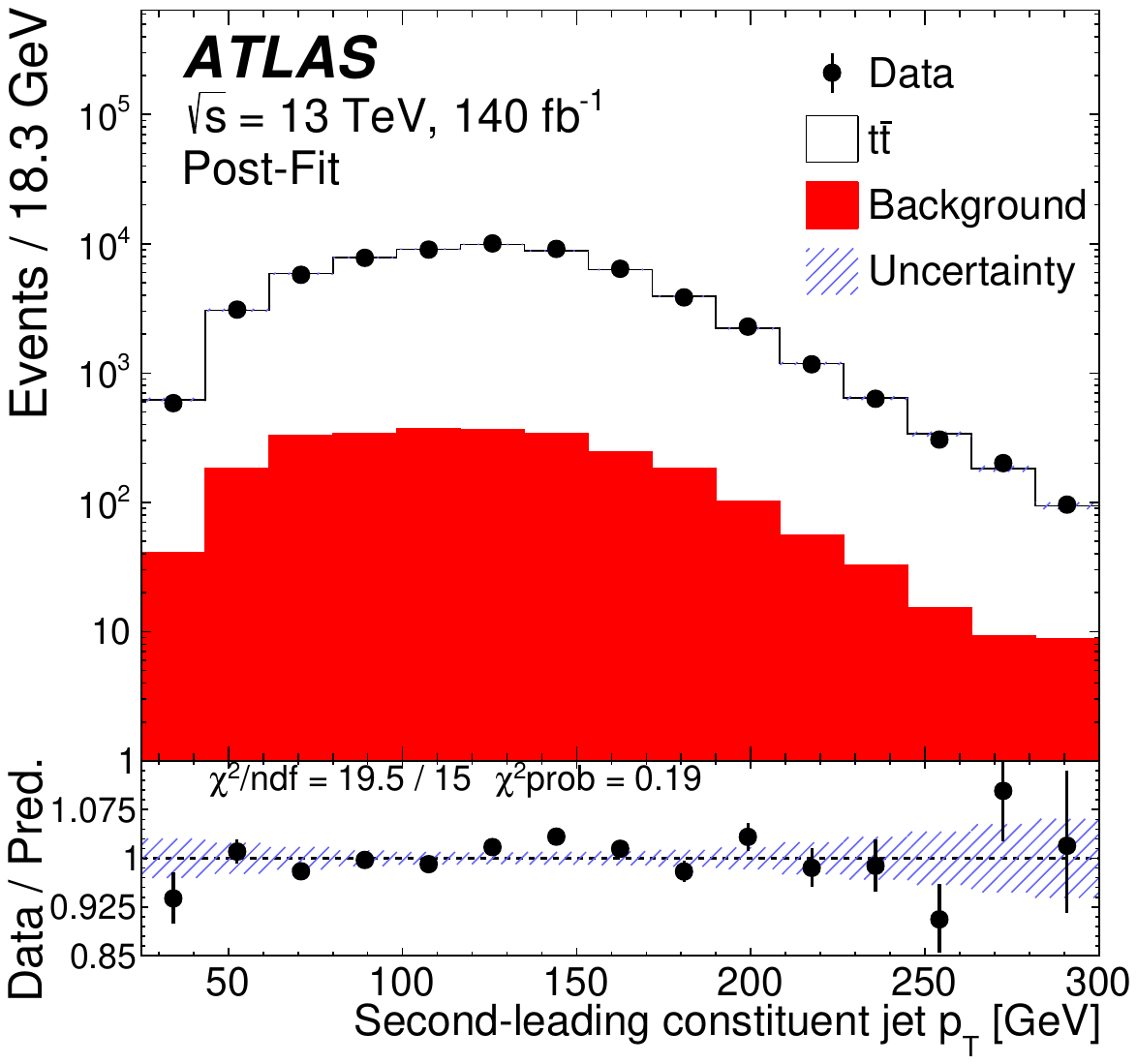}}
\subfloat[]{\includegraphics[width=0.32\textwidth]{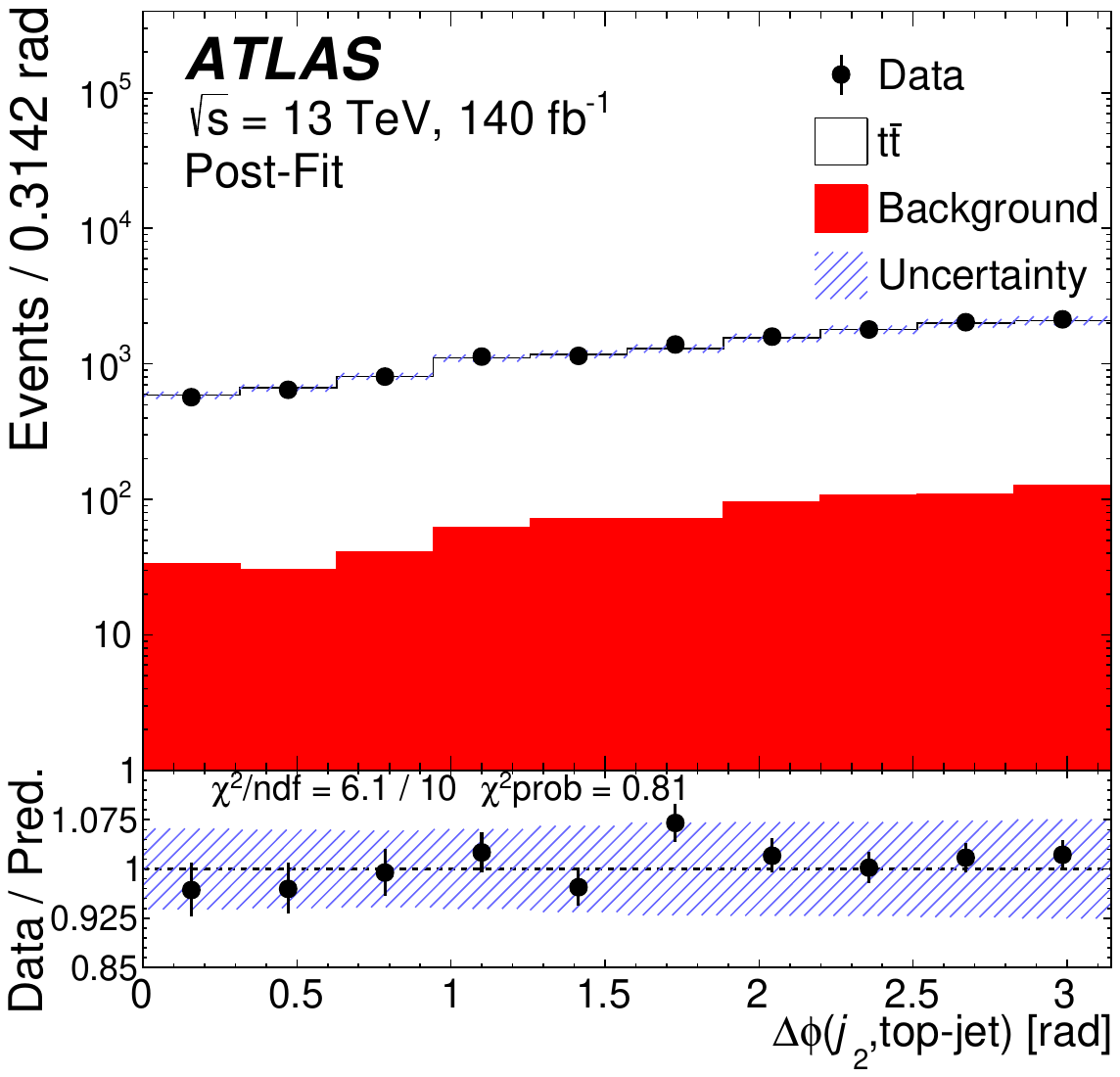}}
\subfloat[]{\includegraphics[width=0.32\textwidth]{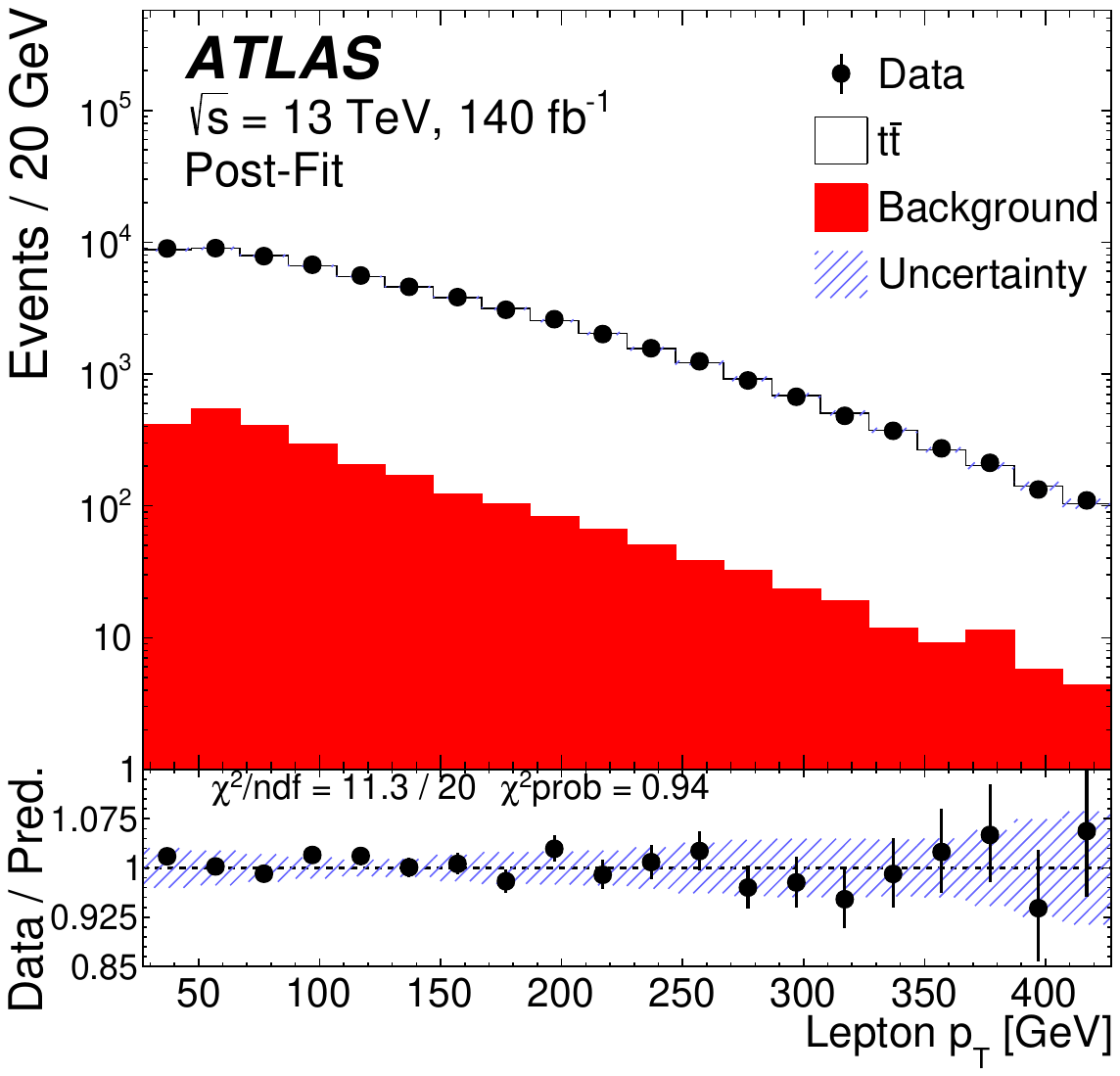}}
\caption{The distributions of (a) the \pt of the second-leading constituent inside the top-jet, (b) the $\Delta\phi$ between the second-leading additional jet and top-jet (for events including at least two additional jets), and (c) the selected lepton \pt are compared with the prediction of the post-fit model. These distributions do not include the overflow of events that can satisfy the event selection of Section~\ref{sec:eventSel}. The uncertainty band includes the effect of the systematic uncertainties, including the post-fit correlations and constraints. The bottom panel reflects the agreement between the data and total prediction, where the error bars show the statistical uncertainty of the data. Values of $\chi^2$ and the number of degrees of freedom (ndf) are provided for each distribution.}
\label{fig:controlRegions}
\end{figure}

\begin{figure}[!htb]
\centering
\subfloat[]{\includegraphics[width=0.4\textwidth]{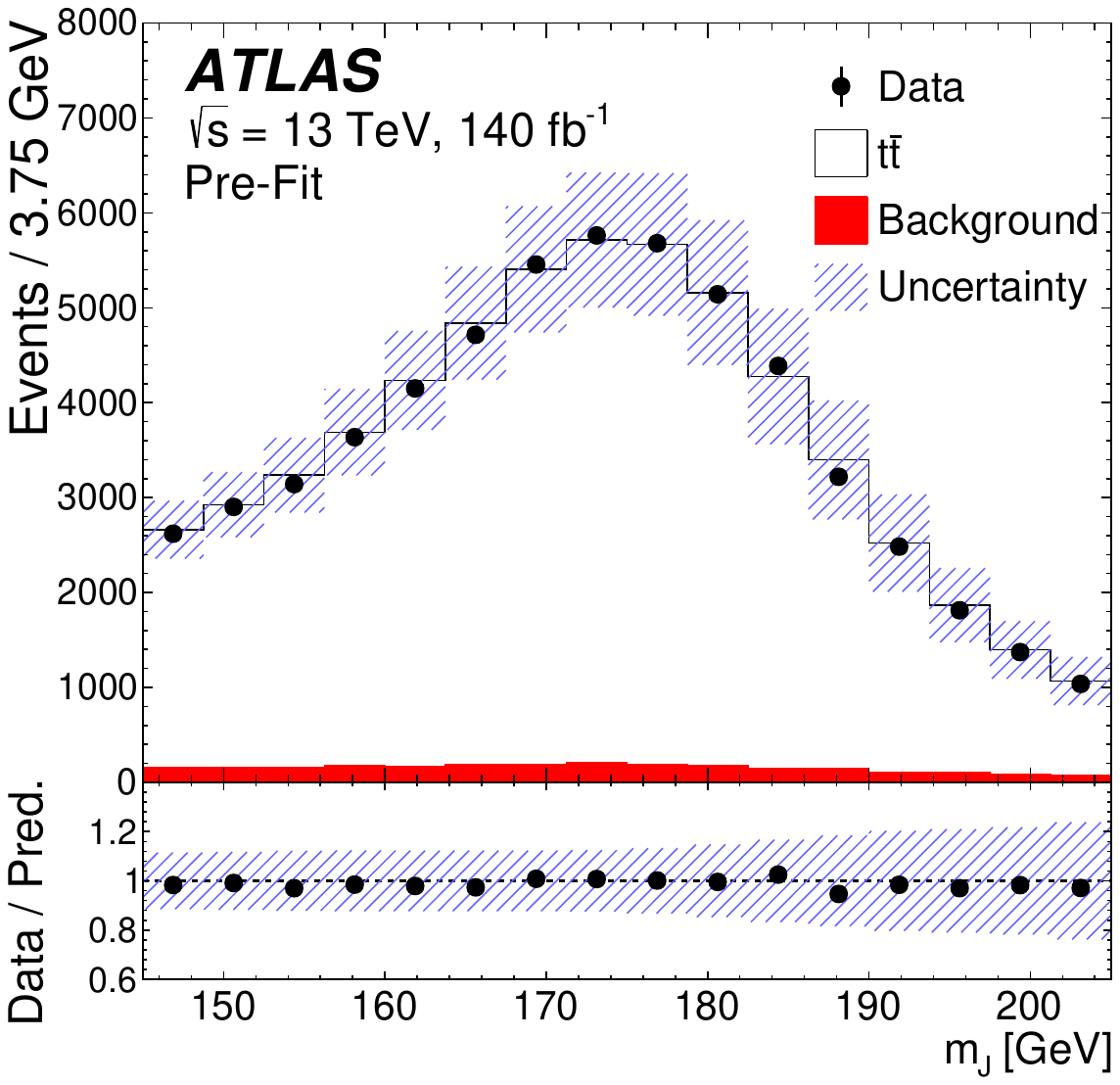}}
\subfloat[]{\includegraphics[width=0.4\textwidth]{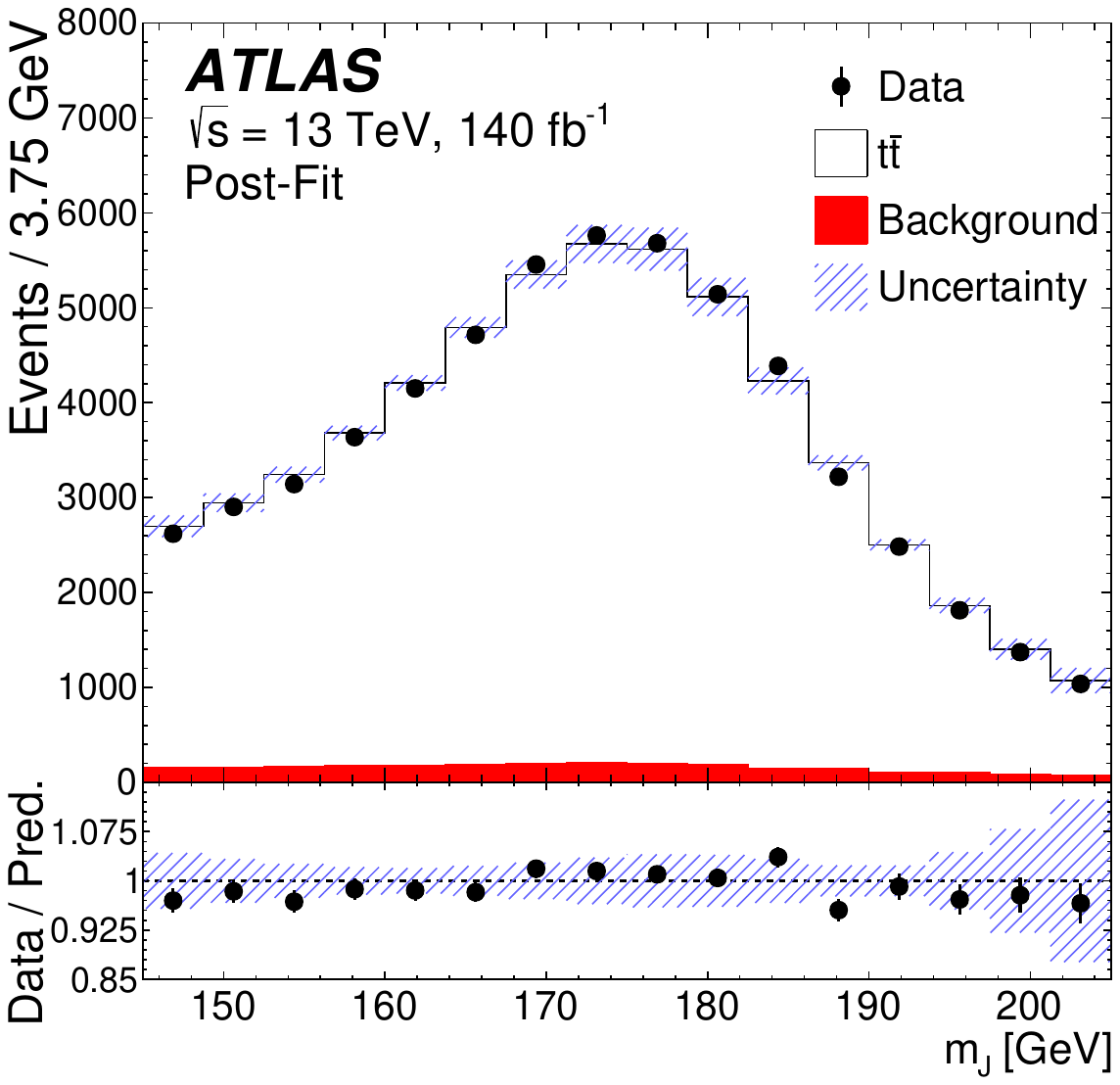}}
\caption{
The (a) pre-fit and (b) post-fit expectations for the \mtopjet distribution are compared with the data.
The fit does not contain the distribution of \mtopjet, but only the value of \meanmtopjet. The distribution is shown to illustrate the agreement between data and prediction.
For the pre-fit case, the $\ttbar$\ sample corresponds to $\mt=172.5$~\GeV and is scaled to $\muttbar=0.856$, which is the ratio of data to theory found in the previous ATLAS cross-section measurement~\cite{TOPQ-2019-23}.
There is no uncertainty in the pre-fit value of \muttbar or \mt.
The uncertainty band represents the total uncertainty in the
model and is significantly reduced after the fit.
The bottom panel reflects the agreement between the data and total prediction, where the error bars show the statistical uncertainty of the data. This includes the systematic and MC statistical uncertainties and for the post-fit case it also includes the uncertainty in  \muttbar~and \mt.
}
\label{fig:finalobs_mJ}
\end{figure}

To validate the robustness of the fit,
the constrained model is propagated to control observables that are not strongly correlated with $\mtopjet$, $\mWreco$~or $\mR$.
Three example distributions, the \pt of the second-leading constituent inside the top-jet, the $\Delta\phi$ between the second-leading additional jet and top-jet (for events including at least two additional jets), and the \pt of the selected lepton, are shown in Figure~\ref{fig:controlRegions}. The \pt of the lepton and the second-leading constituent inside the top-jet probe the modelling of the semi-leptonically decaying top quark and the top-jet, respectively, and the $\Delta\phi$ variable probes the modelling of the parton shower by considering the second additional jet.
A $\chi^2$~test is performed between the data and the model for each observable, where the covariance matrix includes the effect of the constrained systematic uncertainties.
Good agreement is observed between the constrained model and the data; all tested observables have $p$-values greater than $0.19$.
The stability of the result is tested by splitting the data used in the \meanmtopjet~observable into two in three different ways: according to the data taking period (2015--2017, 2018), the lepton flavour or the number
of constituent jets inside the top-jet. In each case, a fit to data is performed with two \mt\ parameters and the fitted values are found to be consistent within one standard deviation, accounting for correlations in the statistical test.
%


%
%
%

The mean of the top-jet mass in data is observed to be $\meanMtopData$~\GeV~and the profile-likelihood fit
gives $\mt = \mtResult \pm \totalUncData$~\GeV.
To visualise the agreement between data and expectation, the fit model is propagated to the binned distribution of $\mtopjet$ and the pre- and post-fit predictions are compared with the data in Figure~\ref{fig:finalobs_mJ}.
The total uncertainty is slightly larger than expected ($\totalUncAsimov$~\GeV) mainly due to the observed number of events being smaller than the expected number.
The breakdown of the uncertainties in different categories is shown
in Table~\ref{table:uncertaintyBreakdown3D}, where the contribution of each systematic uncertainty
is extracted from the covariance matrix of the fit~\cite{Pinto:2023yob} and the total uncertainty per-category is the sum in quadrature of the effect of each uncertainty in the category.
The statistical uncertainty is then calculated from the quadratic difference between the total uncertainty and the total systematic uncertainty (which itself is the sum in quadrature of every systematic uncertainty).
The contributions to the total systematic uncertainty from experimental and theoretical sources are quite similar (0.36 and 0.28~\GeV, respectively)
and the uncertainty category with the largest impact is the JES\@.
The flavour uncertainties for the JES
are reduced compared with previous ATLAS measurements thanks to the improved uncertainty treatment detailed in Refs.~\cite{STDM-2020-04,ATL-PHYS-PUB-2022-021}.
The impact of the most important individual systematic uncertainties is shown in Figure~\ref{fig:np}.
Several aspects of the simulation of \ttbar\ events are important systematic uncertainties,
including the CR model, the parton shower and hadronisation model and the modelling of additional radiation.

%
\begin{table}[htb]
\caption{Grouped breakdown of the uncertainty sources contributing to the total uncertainty in \mt. Each group's contribution is equal to the sum in quadrature of the individual uncertainty contributions. `Radiation' refers to the sources relating to the modelling of ISR and FSR for \ttbar\ events. Those associated with ISR are the setting of \hdamp and the use of the Var3c eigentune of the \Pythia[8] A14 tune. `Higher-order corrections' refers to sources relating to the \muF and \muR scales and the NNLO reweighting. Systematic uncertainties related to the modelling of background processes are contained within the `Background modelling' group.}\label{table:uncertaintyBreakdown3D}
\centering
\setlength\extrarowheight{-2pt}
\resizebox{0.5\columnwidth}{!}{%
\begin{tabular}{lr}
\toprule
Source & Uncertainty $[ \GeV ]$ \\
\midrule
JES & $\pm$ 0.29 \\
Radiation (ISR and FSR) & $\pm$ 0.17 \\
Colour reconnection (CR1 and CR2) & $\pm$ 0.15 \\
JES heavy flavour & $\pm$ 0.14 \\
Parton shower and hadronisation model & $\pm$ 0.14 \\
JER & $\pm$ 0.10 \\
MC statistics & $\pm$ 0.08 \\
Underlying event & $\pm$ 0.08 \\
Recoil & $\pm$ 0.07 \\
Fit closure & $\pm$ 0.07 \\
Background modelling & $\pm$ 0.05 \\
Matrix element matching (\pthard~=~1) & $\pm$ 0.04 \\
$b$-tagging & $\pm$ 0.04 \\
Higher-order corrections & $\pm$ 0.02 \\
\met & $\pm$ 0.02 \\
Pileup & $\pm$ 0.01 \\
JVT & $\pm$ 0.01 \\
PDF & $\pm$ 0.01 \\
Leptons & $\pm$ 0.01 \\
Luminosity & < 0.01 \\
\midrule
Total statistical & $\pm$ 0.27 \\
Total systematic & $\pm$ 0.46 \\
\midrule
Total & $\pm$ 0.53 \\

\bottomrule
\end{tabular}%
}%
\end{table}


A comparison of this result with other $\mt$~measurements is shown in Figure~\ref{fig:summary}.
This new measurement is found to be in excellent agreement with previous ATLAS measurements, the ATLAS+CMS Run 1 combination~\cite{TOPQ-2019-13} and the CMS measurement using boosted top quarks~\cite{CMS-TOP-21-012},
while it is $1.2$~\GeV higher than the most precise CMS measurement performed with Run 2 data~\cite{CMS-TOP-20-008}.
Precisely quantifying the level of agreement between the results displayed in Figure~\ref{fig:summary}~requires a detailed estimate of the correlation
between each pair of measurements, which is beyond the scope of this Letter.

\begin{figure}[htpb]
\centering
\subfloat[]{\includegraphics[width=0.48\textwidth]{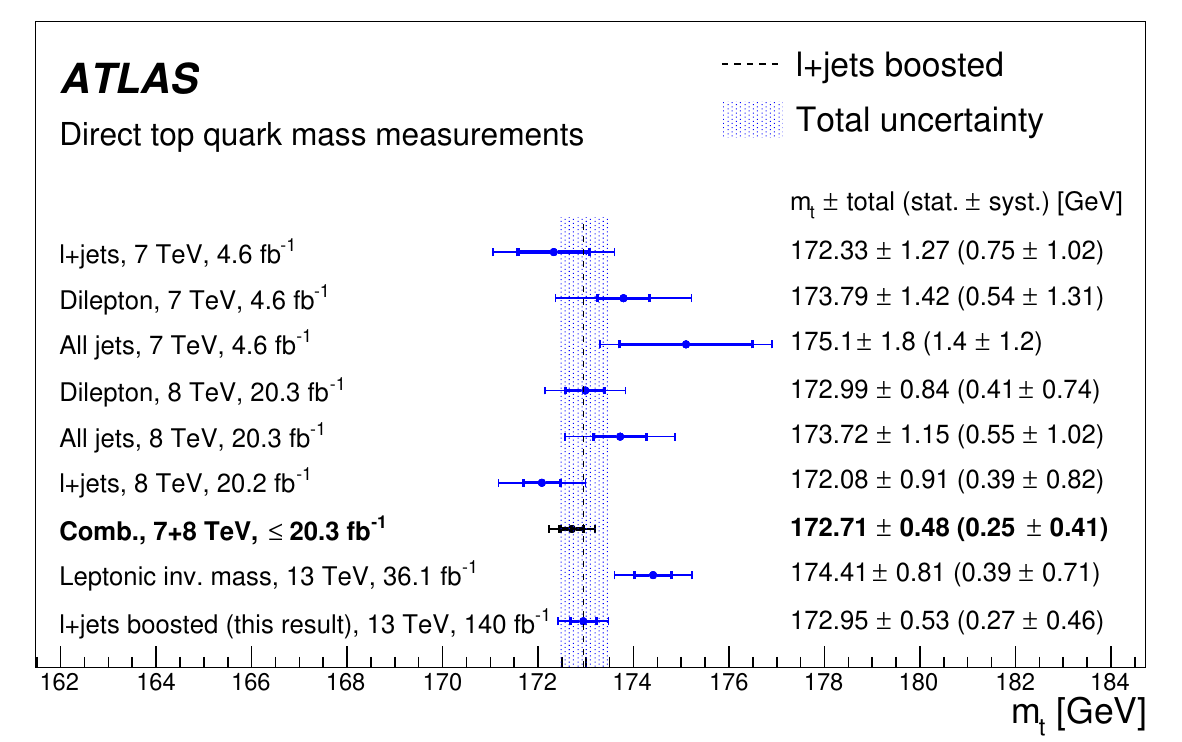}}
\subfloat[]{\includegraphics[width=0.48\textwidth]{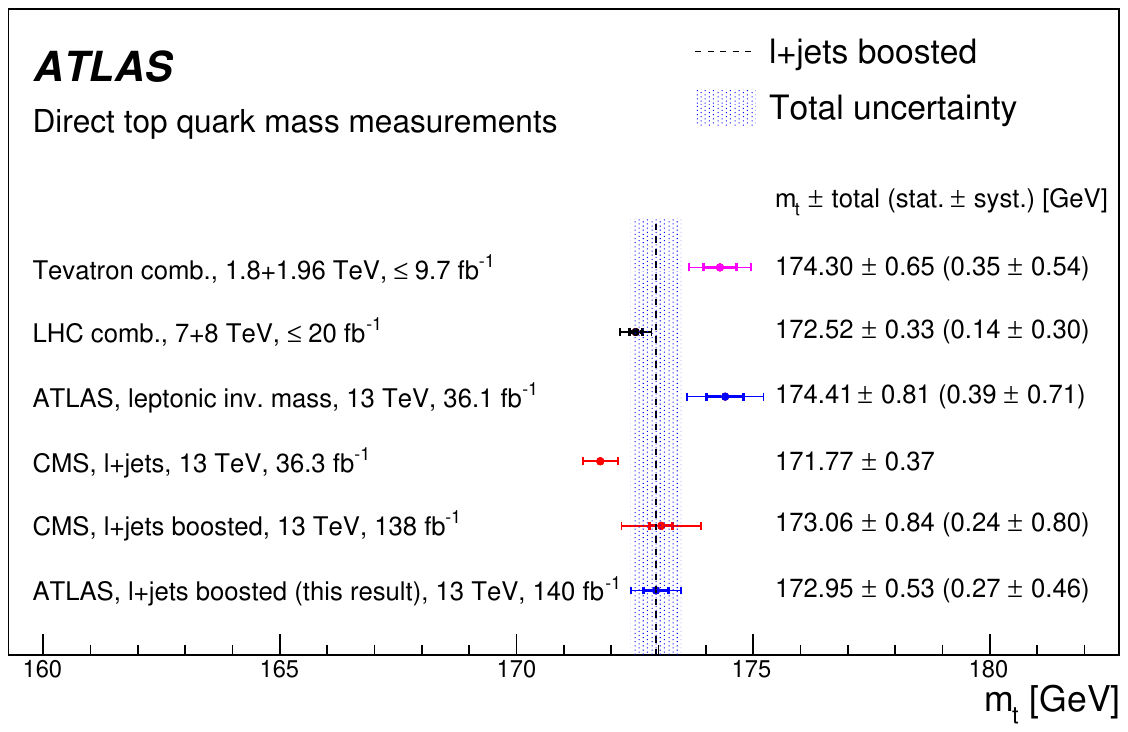}}
\caption{The measurement of the top quark mass in this Letter is compared with (a) other ATLAS measurements and (b) other precise direct determinations of \mt.
The ATLAS combination displayed in bold in (a) is a combination of all the results listed above it.
The dashed line and
shaded band represent the central value and uncertainty of this measurement, which is also shown as the final measurement point in each plot.
}
\label{fig:summary}
\end{figure}


\FloatBarrier
\section{Conclusion}

The top quark mass is measured using $\ttbar$\ events in the lepton-plus-jets decay channel where the hadronically decaying top quark has high transverse momentum. The dataset used corresponds to $140$~\ifb of $pp$~collisions at $13$~\TeV recorded with the ATLAS detector at the LHC.
The boosted hadronically decaying top quark is reconstructed as a large-radius jet and the mean of the invariant mass of the jet is chosen as the observable with high sensitivity to the top quark mass.
Additional observables, $\mWreco$~and $\mR$, are included into the likelihood fit to reduce the impact of the systematic uncertainties, in particular those associated
with the jet energy scale and the modelling of the recoil in the top quark decay.
The top quark mass is measured to be $\mt = \mtResult \pm \totalUncData$~\GeV,
which represents a significant improvement on all previous individual ATLAS measurements and is in good agreement with other top quark mass measurements.
The measurement also achieves an improved precision compared to previous measurements in events with boosted top quarks made by the CMS Collaboration.
The largest systematic uncertainties stem from the understanding of the jet energy scale
and the modelling of $\ttbar$~events.
The measurement has a non-negligible statistical uncertainty, which indicates that future measurements in the boosted phase space
with larger LHC datasets could reach an improved precision.


%
%
%
%
%
\section*{Acknowledgements}
%

%

%
%

%
%

We thank CERN for the very successful operation of the LHC and its injectors, as well as the support staff at
CERN and at our institutions worldwide without whom ATLAS could not be operated efficiently.

The crucial computing support from all WLCG partners is acknowledged gratefully, in particular from CERN, the ATLAS Tier-1 facilities at TRIUMF/SFU (Canada), NDGF (Denmark, Norway, Sweden), CC-IN2P3 (France), KIT/GridKA (Germany), INFN-CNAF (Italy), NL-T1 (Netherlands), PIC (Spain), RAL (UK) and BNL (USA), the Tier-2 facilities worldwide and large non-WLCG resource providers. Major contributors of computing resources are listed in Ref.~\cite{ATL-SOFT-PUB-2025-001}.

We gratefully acknowledge the support of ANPCyT, Argentina; YerPhI, Armenia; ARC, Australia; BMWFW and FWF, Austria; ANAS, Azerbaijan; CNPq and FAPESP, Brazil; NSERC, NRC and CFI, Canada; CERN; ANID, Chile; CAS, MOST and NSFC, China; Minciencias, Colombia; MEYS CR, Czech Republic; DNRF and DNSRC, Denmark; IN2P3-CNRS and CEA-DRF/IRFU, France; SRNSFG, Georgia; BMBF, HGF and MPG, Germany; GSRI, Greece; RGC and Hong Kong SAR, China; ICHEP and Academy of Sciences and Humanities, Israel; INFN, Italy; MEXT and JSPS, Japan; CNRST, Morocco; NWO, Netherlands; RCN, Norway; MNiSW, Poland; FCT, Portugal; MNE/IFA, Romania; MSTDI, Serbia; MSSR, Slovakia; ARIS and MVZI, Slovenia; DSI/NRF, South Africa; MICIU/AEI, Spain; SRC and Wallenberg Foundation, Sweden; SERI, SNSF and Cantons of Bern and Geneva, Switzerland; NSTC, Taipei; TENMAK, T\"urkiye; STFC/UKRI, United Kingdom; DOE and NSF, United States of America.

Individual groups and members have received support from BCKDF, CANARIE, CRC and DRAC, Canada; CERN-CZ, FORTE and PRIMUS, Czech Republic; COST, ERC, ERDF, Horizon 2020, ICSC-NextGenerationEU and Marie Sk{\l}odowska-Curie Actions, European Union; Investissements d'Avenir Labex, Investissements d'Avenir Idex and ANR, France; DFG and AvH Foundation, Germany; Herakleitos, Thales and Aristeia programmes co-financed by EU-ESF and the Greek NSRF, Greece; BSF-NSF and MINERVA, Israel; NCN and NAWA, Poland; La Caixa Banking Foundation, CERCA Programme Generalitat de Catalunya and PROMETEO and GenT Programmes Generalitat Valenciana, Spain; G\"{o}ran Gustafssons Stiftelse, Sweden; The Royal Society and Leverhulme Trust, United Kingdom.

In addition, individual members wish to acknowledge support from Armenia: Yerevan Physics Institute (FAPERJ); CERN: European Organization for Nuclear Research (CERN DOCT); Chile: Agencia Nacional de Investigaci\'on y Desarrollo (FONDECYT 1230812, FONDECYT 1230987, FONDECYT 1240864); China: Chinese Ministry of Science and Technology (MOST-2023YFA1605700, MOST-2023YFA1609300), National Natural Science Foundation of China (NSFC - 12175119, NSFC 12275265, NSFC-12075060); Czech Republic: Czech Science Foundation (GACR - 24-11373S), Ministry of Education Youth and Sports (ERC-CZ-LL2327, FORTE CZ.02.01.01/00/22\_008/0004632), PRIMUS Research Programme (PRIMUS/21/SCI/017); EU: H2020 European Research Council (ERC - 101002463); European Union: European Research Council (ERC - 948254, ERC 101089007, ERC, BARD, 101116429), Horizon 2020 Framework Programme (MUCCA - CHIST-ERA-19-XAI-00), European Union, Future Artificial Intelligence Research (FAIR-NextGenerationEU PE00000013), Italian Center for High Performance Computing, Big Data and Quantum Computing (ICSC, NextGenerationEU); France: Agence Nationale de la Recherche (ANR-20-CE31-0013, ANR-21-CE31-0013, ANR-21-CE31-0022, ANR-22-EDIR-0002); Germany: Baden-Württemberg Stiftung (BW Stiftung-Postdoc Eliteprogramme), Deutsche Forschungsgemeinschaft (DFG - 469666862, DFG - CR 312/5-2); Italy: Istituto Nazionale di Fisica Nucleare (ICSC, NextGenerationEU), Ministero dell'Università e della Ricerca (NextGenEU PRIN20223N7F8K M4C2.1.1); Japan: Japan Society for the Promotion of Science (JSPS KAKENHI JP22H01227, JSPS KAKENHI JP22H04944, JSPS KAKENHI JP22KK0227, JSPS KAKENHI JP23KK0245); Norway: Research Council of Norway (RCN-314472); Poland: Ministry of Science and Higher Education (IDUB AGH, POB8, D4 no 9722), Polish National Science Centre (NCN 2021/42/E/ST2/00350, NCN OPUS 2023/51/B/ST2/02507, NCN OPUS nr 2022/47/B/ST2/03059, NCN UMO-2019/34/E/ST2/00393, UMO-2020/37/B/ST2/01043, UMO-2022/47/O/ST2/00148, UMO-2023/49/B/ST2/04085, UMO-2023/51/B/ST2/00920); Portugal: Foundation for Science and Technology (FCT); Spain: Generalitat Valenciana (Artemisa, FEDER, IDIFEDER/2018/048), Ministry of Science and Innovation (MCIN \& NextGenEU PCI2022-135018-2, MICIN \& FEDER PID2021-125273NB, RYC2019-028510-I, RYC2020-030254-I, RYC2021-031273-I, RYC2022-038164-I); Sweden: Carl Trygger Foundation (Carl Trygger Foundation CTS 22:2312), Swedish Research Council (Swedish Research Council 2023-04654, VR 2018-00482, VR 2021-03651, VR 2022-03845, VR 2022-04683, VR 2023-03403), Knut and Alice Wallenberg Foundation (KAW 2018.0458, KAW 2019.0447, KAW 2022.0358); Switzerland: Swiss National Science Foundation (SNSF - PCEFP2\_194658); United Kingdom: Leverhulme Trust (Leverhulme Trust RPG-2020-004), Royal Society (NIF-R1-231091); United States of America: U.S. Department of Energy (ECA DE-AC02-76SF00515), Neubauer Family Foundation.

%
%


%
%
%
%
%
%
%
%
%

%
%
%
\printbibliography
\clearpage
 
\begin{flushleft}
\hypersetup{urlcolor=black}
{\Large The ATLAS Collaboration}

\bigskip

\AtlasOrcid[0000-0002-6665-4934]{G.~Aad}$^\textrm{\scriptsize 104}$,
\AtlasOrcid[0000-0001-7616-1554]{E.~Aakvaag}$^\textrm{\scriptsize 17}$,
\AtlasOrcid[0000-0002-5888-2734]{B.~Abbott}$^\textrm{\scriptsize 123}$,
\AtlasOrcid[0000-0002-0287-5869]{S.~Abdelhameed}$^\textrm{\scriptsize 119a}$,
\AtlasOrcid[0000-0002-1002-1652]{K.~Abeling}$^\textrm{\scriptsize 55}$,
\AtlasOrcid[0000-0001-5763-2760]{N.J.~Abicht}$^\textrm{\scriptsize 49}$,
\AtlasOrcid[0000-0002-8496-9294]{S.H.~Abidi}$^\textrm{\scriptsize 30}$,
\AtlasOrcid[0009-0003-6578-220X]{M.~Aboelela}$^\textrm{\scriptsize 45}$,
\AtlasOrcid[0000-0002-9987-2292]{A.~Aboulhorma}$^\textrm{\scriptsize 36e}$,
\AtlasOrcid[0000-0001-5329-6640]{H.~Abramowicz}$^\textrm{\scriptsize 157}$
% ...
\bigskip
\\

% ...
$^{17}$Department for Physics and Technology, University of Bergen, Bergen; Norway.\\
% ...
$^{30}$Physics Department, Brookhaven National Laboratory, Upton NY; United States of America.\\
% ...
$^{36}$$^{(a)}$Facult\'e des Sciences Ain Chock, Universit\'e Hassan II de Casablanca;$^{(b)}$Facult\'{e} des Sciences, Universit\'{e} Ibn-Tofail, K\'{e}nitra;$^{(c)}$Facult\'e des Sciences Semlalia, Universit\'e Cadi Ayyad, LPHEA-Marrakech;$^{(d)}$LPMR, Facult\'e des Sciences, Universit\'e Mohamed Premier, Oujda;$^{(e)}$Facult\'e des sciences, Universit\'e Mohammed V, Rabat;$^{(f)}$Institute of Applied Physics, Mohammed VI Polytechnic University, Ben Guerir; Morocco.\\
% ...
$^{45}$Physics Department, Southern Methodist University, Dallas TX; United States of America.\\
% ...
$^{49}$Fakult\"{a}t Physik , Technische Universit{\"a}t Dortmund, Dortmund; Germany.\\
% ...
$^{55}$II. Physikalisches Institut, Georg-August-Universit\"{a}t G\"ottingen, G\"ottingen; Germany.\\
% ...
$^{104}$CPPM, Aix-Marseille Universit\'e, CNRS/IN2P3, Marseille; France.\\
% ...
$^{119}$$^{(a)}$New York University Abu Dhabi, Abu Dhabi;$^{(b)}$United Arab Emirates University, Al Ain; United Arab Emirates.\\
% ...
$^{123}$Homer L. Dodge Department of Physics and Astronomy, University of Oklahoma, Norman OK; United States of America.\\
% ...
$^{157}$Raymond and Beverly Sackler School of Physics and Astronomy, Tel Aviv University, Tel Aviv; Israel.\\
% ...
$^{a}$ Also at Affiliated with an institute formerly covered by a cooperation agreement with CERN.\\
$^{b}$ Also at An-Najah National University, Nablus; Palestine.\\
$^{c}$ Also at Borough of Manhattan Community College, City University of New York, New York NY; United States of America.\\
$^{d}$ Also at Center for High Energy Physics, Peking University; China.\\
$^{e}$ Also at Center for Interdisciplinary Research and Innovation (CIRI-AUTH), Thessaloniki; Greece.\\
$^{f}$ Also at Centre of Physics of the Universities of Minho and Porto (CF-UM-UP); Portugal.\\
% ...

\end{flushleft}


%

%
%

%

%
%
%
%

\end{document}